\documentclass[preprint,5p, twocolumn, 12pt]{elsarticle}
\usepackage{amssymb}
\usepackage[fleqn]{amsmath}
\usepackage{mathptmx}
\usepackage{xurl}
\usepackage[T1]{fontenc}
\usepackage{xstring}
\usepackage{lipsum}
\usepackage{graphicx} 
\usepackage{subcaption}
\usepackage{tabularx}
\usepackage{boldline}
\usepackage{float}
\usepackage{bm}
\usepackage{csquotes}
\usepackage{algorithm}
\usepackage{algpseudocode}
\usepackage{newpxtext} 
\usepackage{newpxmath} 
\usepackage[labelfont=bf,justification=raggedright,singlelinecheck=false]{caption}
\usepackage[font=footnotesize,labelfont=bf]{caption}
\usepackage[font=footnotesize,labelfont=bf]{subcaption}
\usepackage{hyperref}
\hypersetup{
	colorlinks   = true,
	citecolor    = blue,
	linkcolor=blue  
}
\usepackage{cleveref}

\makeatletter
\newtoks\@eadauthorshort
\def\@author#1#2{\g@addto@macro\elsauthors{\normalsize%
		\def\baselinestretch{1}%
		\upshape\authorsep#1\unskip\textsuperscript{%
			\ifx\@fnmark\@empty\else\unskip\sep\@fnmark\let\sep=,\fi
			\ifx\@corref\@empty\else\unskip\sep\@corref\let\sep=,\fi
		}%
		\def\authorsep{\unskip,\space}%
		\global\let\@fnmark\@empty
		\global\let\sep\@empty}%
	\@eadauthor={#1}
	\@eadauthorshort={#2}
}
\def\@@author[#1]#2#3{\g@addto@macro\elsauthors{%
		\def\baselinestretch{1}%
		\authorsep#2\unskip\textsuperscript{
			\@for\@@affmark:=#1\do{%
				\edef\affnum{\@ifundefined{X@\@@affmark}{1}{\elsRef{\@@affmark}}}%
				\unskip\sep\affnum\let\sep=,}%
			\ifx\@fnmark\@empty\else\unskip\sep\@fnmark\let\sep=,\fi
			\ifx\@corref\@empty\else\unskip\sep\@corref\let\sep=,\fi
		}%
		\def\authorsep{\unskip,\space}%
		\global\let\sep\@empty\global\let\@corref\@empty
		\global\let\@fnmark\@empty}%
	\@eadauthor={#2}%
	\@eadauthorshort={#3}%
}
\gdef\@ead#1{\bgroup\def\_{\string\underscorechar\space}%
	\def\{{\string\lbracechar\space}%
	\def~{\hashchar\space}%
	\def\}{\string\rbracechar\space}%
	\edef\tmpA{\the\@eadauthor}
	\edef\tmpB{\the\@eadauthorshort}
	\immediate\write\@auxout{\string\emailauthor
		{#1}{\expandafter\strip@prefix\meaning\tmpA}{\expandafter\strip@prefix\meaning\tmpB}}%
	\egroup
}
\gdef\emailauthor#1#2#3{\stepcounter{ead}%
	\g@addto@macro\@elseads{\raggedright%
		\let\corref\@gobble
		\eadsep\texttt{#1} (\ifemailshortauthor #3\else#2\fi)\def\eadsep{\unskip,\space}}%
}
\newif\ifemailshortauthor

\providecommand{\doi}[1]{%
	\begingroup
	\let\bibinfo\@secondoftwo
	\urlstyle{rm}%
	\href{http://dx.doi.org/#1}{%
		doi:\discretionary{}{}{}%
		\nolinkurl{#1}%
	}%
	\endgroup
}
\makeatother

\AtBeginDocument{
	\let\oldref\ref
	\renewcommand{\ref}[1]{\IfBeginWith{#1}{fig:}%
		{{\color{blue}Figure~\oldref{#1}}}%
		{\IfBeginWith{#1}{tab:}{{\color{blue}Table~\oldref{#1}}}{Unsupported ref start}}}%
}
\makeatother

\emailshortauthortrue

\begin{document}
	\begin{frontmatter}
		\title{On the Application of Fractional Order Derivatives for Characterizing Brain White Matter Viscoelasticity}
		\author[1]{Parameshwaran Pasupathy}{P Pasupathy}
		
		\author[2]{John G Georgiadis}{JG Georgiadis}
		
		\author[1]{Assimina A Pelegri\corref{correspondingauthor}}{AA Pelegri}
		\cortext[correspondingauthor]{Corresponding author}
		\ead{pelegri@rutgers.edu}
		
		\address[1]{Department of Mechanical and Aerospace Engineering, Rutgers, The State University of New Jersey, Piscataway, NJ 08854, USA}
		\address[2]{Department of Biomedical Engineering, Illinois Tech, Chicago, IL 60616, USA}
		\begin{abstract}
			
			\noindent \textbf{Background:} Conventional viscoelastic characterization of brain white matter (BWM), typically accomplished within the classical framework of springs and dashpots expressed as a Prony series, remains a purely empirical representation that is difficult to interpret physically. A growing body of research suggests that the viscoelastic response of BWM can be described by power-law behavior. A power-law model in the frequency domain, under the assumptions of linear viscoelasticity and causality, yields a fractional viscoelastic model in the time domain.
			
			\noindent \textbf{Methods:}  We develop a fractional viscoelastic model of the axon and extracellular matrix (ECM). which is implemented in a Fortran \texttt{VUMAT} subroutine. A biphasic finite element model of hexagonally packed representative volume elements (RVEs) of axons in ECM, subjected to periodic and quasi-static displacement-controlled boundary conditions, is built in Abaqus. An optimization algorithm is then applied to solve the inverse problem and extract the homogenized properties of the biphasic model.
			
			\noindent \textbf{Results:} The model predicts that the spring-pot coefficient, which determines the solid-fluid behavior, varies linearly with volume fraction along the fiber direction, but follows a bi-logistic function along the transverse normal and shear directions. The power-law exponent, which encodes information about the underlying tissue architecture, exhibits nonlinear trends in all loading directions: a saturating exponential along the fiber direction and bi-logistic variation in the transverse and shear directions. The nonlinear variation of the parameters reveals two distinct stiffening stages: a lower rate at low axon volume fractions, followed by a higher rate as increased axonal content reinforces the RVE.
			
			\noindent \textbf{Significance:} To the best of our knowledge, this study is the first to propose and implement a 3D fractional viscoelastic finite element model of the corpus callosum of BWM in the time domain. This framework, by incorporating axonal configuration and mechanical properties,	 provides a succinct, physically meaningful description of BWM’s mechanical response, which is difficult to capture with Prony series models. The results reveal the nonlinear variation of material parameters with axon volume fraction, the directional dependence of BWM mechanics, and the complex interplay among microstructural elements—all of which are crucial for understanding neuropathological changes associated with disease and aging. In addition, we demonstrate an efficient, thread-safe numerical implementation of the fractional viscoelastic model that achieves significantly faster performance compared to existing implementations.
			
		\end{abstract}
		
		
	\end{frontmatter}

	\section{Introduction}
	Constituting approximately 50 percent of the human brain, white matter (BWM) plays a critical role in maintaining brain health, making it a key area related to diagnostics and therapeutics targeting the brain parenchyma. BWM contains billions of myelinated axons organized in 21 major tracts whose primary function is to facilitate communication between different brain regions \cite{bullock_taxonomy_2022}. Unlike other organs, such as the heart, for which robust prediction models backed by experiments and large scientific communities exist, mechanical characterization of the brain---particularly BWM---remains challenging. 
	
	Relating BWM response to external mechanical excitation requires a robust integration of biomechanical experiments and microstructural modeling. The above logical inference, however, belies the enormity of the task. Firstly, brain mechanics involves disparate scales, i.e., a highly compliant mechanical response under an applied load while being completely enclosed inside a hard skull. This makes the brain parenchyma hard to probe. Further complications in experimentation arise from the fact that the mechanical properties of brain tissue are highly multiphasic and region-specific. The large scatter of data and variations in results between \textsl{in vitro} experimental results and \textsl{in vivo} measurements are a testament to this fact, all before considering diverse factors such as gender, age, disease, past injury/trauma \cite{chatelin_fifty_2010, brainsci14040401, sack_impact_2009}. 
	
\begin{figure}[h!]
		\begin{subfigure}[b]{\columnwidth}
			\includegraphics[width=\linewidth]{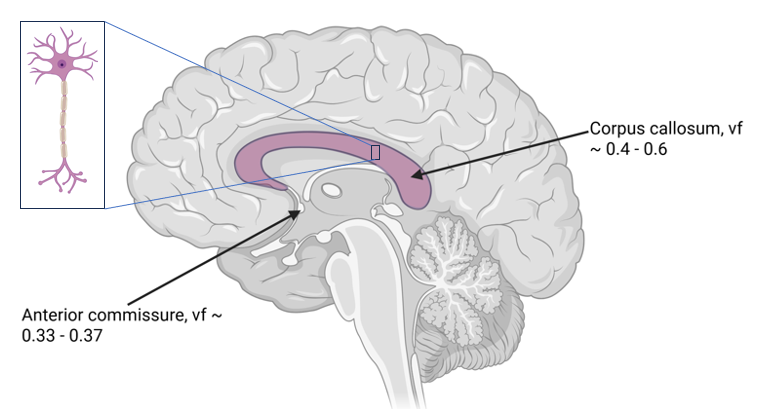}
			\caption{}
			\label{subfig:wm_sagittal_cut}
		\end{subfigure}
		\hfill
		\begin{subfigure}[b]{\columnwidth}
			\includegraphics[width=0.85\linewidth]{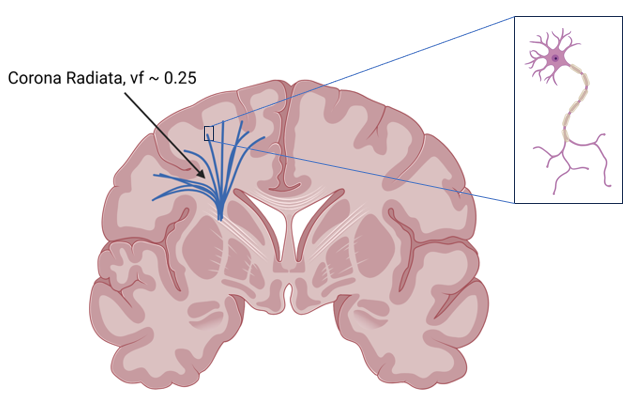}
			\caption{}
			\label{subfig:wm_coronal_cut}
		\end{subfigure} 
		\caption{A schematic representation of the brain examined through different cut planes \cite{eilbes2025neuroanatomical}. (a)  Sagittal cut of the brain depicting the CC of the brain containing aligned myelinated axons with volume fraction (vf): 0.4 - 0.6 \cite{suzuki_estimation_2016} and, anterior commissure, vf: 0.33 - 0.37 \cite{edwards_microstructural_2024}. (b) Coronal cut depicting white matter and corona radiata containing highly tortuous myelinated axons propagating outward with a vf: \textasciitilde  0.25 \cite{huang_high-gradient_2020}. }
		\label{fig:wm_schematic}
	\end{figure}

	Additionally, axon concentration, thickness, distribution and orientation vary widely from region to region making BWM highly heterogenous and anisotropic (\ref{fig:wm_schematic}). For instance, within the corona radiata, the axons that propagate outward are highly tortuous with lower volume fractions, whereas in the corpus callosum (CC), they are aligned and densely packed. Moreover, material characterization of BWM, even under the limit of small strains is difficult owing to its dependence on both magnitude and rate of strain. Quantifying this heterogeneity in BWM is therefore critical for addressing a wide range of questions related to aging, disease onset, and progression \cite{burzynska_correlates_2024}. In light of the above considerations, this study focuses on developing computational models that describe the viscoelastic material response of BWM under mechanical loading, with particular focus on regions with aligned axonal bundles, where such models are most appropriate.
	
	The past two decades have seen significant progress in \textsl{in vivo} and \textsl{ex vivo} mechanical characterization of BWM, with growing evidence indicating that axonal damage is the leading cause of traumatic brain injury (TBI), with excessive tensile strain postulated as the underlying mechanism \cite{arbogast1998material}. Moreover, diffuse axonal injury, which leads to impaired consciousness in patients, occurs due to the shearing of axons \cite{SAIKI2009549}, highlighting the importance of characterizing the mechanical response of myelinated axon bundles in BWM with high fidelity. 
	
	\textsl{In vivo} magnetic resonance imaging (MRI) has allowed the estimation of microstructural parameters related to the spatial distribution of axons in BWM. In particular, diffusion-based MRI techniques have been used to quantify mean axonal diameter, inter-axonal volume fraction, and axon density across healthy young, and aging brains. Studies report that in healthy adults, the mean axon diameter is larger in the body of the CC and smaller in the genu and splenium, whereas the opposite trend was observed for the inter-axonal volume fraction \cite{suzuki_estimation_2016}. In aging brains, a global increase in axon diameter index and a decrease in packing density throughout the CC has been observed with the effect being most pronounced in the genu \cite{fan_age-related_2019}.

	The advent of elastography (MRE), coupled with \textsl{ex vivo} experiments and modeling techniques such as the finite element method, molecular dynamics, and more recently, machine learning, has yielded promising results for \textsl{in vivo} characterization of BWM. Researchers using MRE have observed wide variations in region-specific mechanical properties of BWM compared to global white matter properties \cite{johnson_local_2013}. Moreover, MRE-based studies have documented significant degradation of BWM mechanical properties in patients with multiple sclerosis \cite{WUERFEL20102520}, as well as disability-associated effects that appear independent of disease duration \cite{kiss_low-frequency_2024}. 
	  
	Several computational models of the microstructure of BWM, based on structure–property relationships gleaned from MRI/MRE, have been proposed. These span triphasic unidirectional composite models, consisting of axons, myelin, and extracellular matrix (ECM), which estimate homogenized viscoelastic properties under steady-state oscillatory dynamics \cite{sullivan_sensitivity_2021}, to single-axon models with substructure resolutions under varying strain rates, investigating axonal injury mechanisms \cite{montanino2018}. A number of these studies use Prony series models for viscoelastic material characterization in the time domain. Javid et al. \cite{JAVID2014290} used the genetic algorithm optimization procedure to determine the homogenized Prony series parameters for representative volume elements (RVEs) of axons in the ECM. The objective function was defined using relaxation tests performed on the porcine brain white matter. A microstructure-informed finite element model with an Ogden hyperelastic strain energy density function and a two term Prony series viscoelastic model under cyclic loading developed by Reiter et al. \cite{reiter_modeling_2023} observed that viscoelastic time constants are transferable between porcine and human brain tissue. Li et al. \cite{li_investigation_2021} estimated the linear viscoelastic material properties of grey and white matter of bovine brain tissue through a Prony series model under compression loading. An MRI–based FEM model of a mouse brain coupled with a convolutional neural network (CNN) algorithm for cortical impact simulations was developed by Lai et al. \cite{lai_machine_2020}. The anatomical details were mapped to each FEM element from the corresponding image voxel. A CNN network was used to reconstruct strain-maps from FEM data. Wu et al. \cite{wu_application_2024} developed a multiscale 3D residual neural network algorithm that utilizes a voxelization method to obtain geometry information from 3D RVEs that are based on MRI/MRE data. The architecture information is then encoded in the voxelized location while cross-referencing the RVEs’ material properties. 
	
	Viscoelastic material characterization via Prony series can replicate phenomenological behavior; however, its physical interpretation is often challenging. Even a simple two-parameter Prony series model requires the estimation of five material constants. A number of research studies indicate that the viscoelastic response of BWM can be phenomenologically explained by a power-law behavior. Studies employing multifrequency MRE fitted to a power-law model have characterized age-related alterations in BWM \cite{sack_impact_2009, sack2013structure} and the impact of neuroinflammation on brain viscoelastic properties during the early stages of multiple sclerosis \cite{fehlner_higherresolution_2016}. Kurt et al. \cite{kurt2019optimization} developed a protocol to calculate optimal frequency sets and determine power-law parameters separately for the entire brain, white matter and grey matter. Using ultrasound shear wave elastography, Nicolas et al. \cite{nicolas2018biomechanical} performed \textsl{ex vivo} brain experiments  and determined its mechanical parameters by fitting a power-law model. Kang et al. \cite{kang_viscoelastic_2024} used a power-law model to describe the creep and relaxation profiles of adult porcine white and gray matter under uniaxial compression tests. Mishra and Cleveland \cite{mishra_rheological_2024} conducted low-frequency oscillatory shear tests to measure the rheological properties of porcine organs, including the brain, and developed the Semi-Fractional Kelvin–Voigt model to characterize their mechanical response. 
	
	It is well established that both geometric and dynamic self-similarity, as well as scale invariance, are prominently expressed in the brain \cite{grosu_fractal_2023}. Building on this finding, we posit in the present study that the viscoelastic power-law behavior observed at the tissue level extends across length scales, from the continuum macroscopic level to the microstructural realm of the axons. A viscoelastic power-law model of a spring-pot is applied to the axons and the ECM. The material parameters for the spring-pot are obtained via a logistic regression analysis. A 3D fractional viscoelastic model is then implemented within the finite element framework. Homogenized fractional viscoelastic parameters as a function of the volume fractions of axons embedded in the ECM are finally derived \cite{pasupathy_fractional_2023}.

	\section{Methods}
	
\begin{figure*}[htb]
		\centering
		\includegraphics[width=0.97\textwidth]{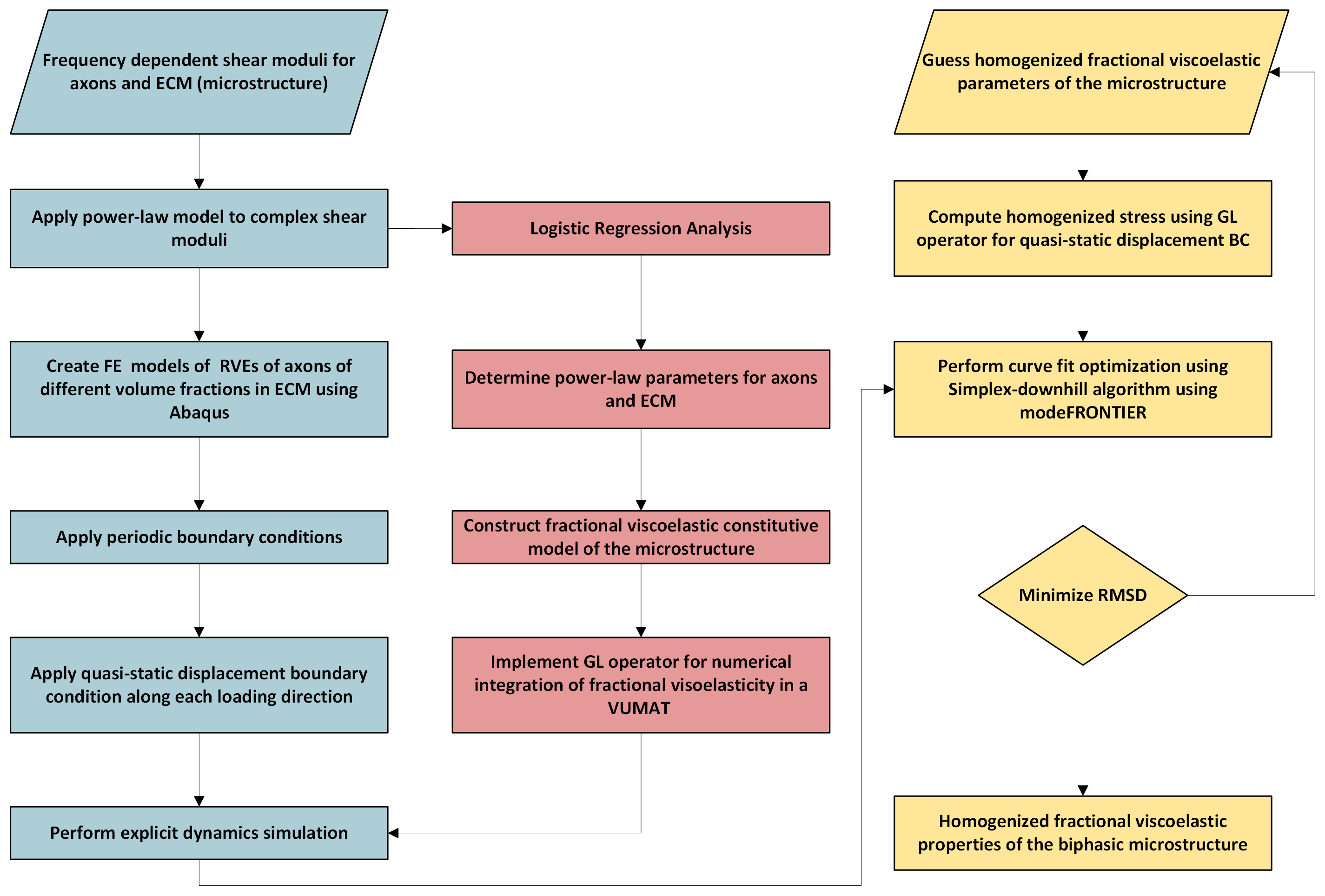}
		\caption{An algorthmic worflow for developing a fractional viscoelastic model of 3D biphasic unit cells representing axon and glia.}\label{fig:Flowchart}
	\end{figure*}

	 \ref{fig:Flowchart} illustrates the workflow of implementing the fractional viscoelastic model for BWM. The remainder of this article essentially delves into each component of the model. \Cref{subsec:regression} describes the logistic regression model for estimating power-law parameters of the microstructure of BWM. The implementation and validation of the fractional viscoelastic model are discussed in \cref{subsec:cmm} --- \cref{subsec:mv}. The results are documented and discussed in \cref{sec:Res} and \cref{sec:Discussion}.
	 
	 \subsection{Logistic Regression Model for Power-Law} 
	 \label{subsec:regression}
	
	The stress-strain response for a viscoelastic material at time \(t\) with the action of stress beginning at time \(\tau\) can be described using a convolution integral of the form
	
	\begin{equation}
	\label{eq:eq_1}
	\sigma(t) = \int_{0}^{t} G(t- \tau) \frac{d \epsilon(\tau)}{d \tau} \hspace{2mm}d\tau,
	\end{equation}

	\noindent and the strain as a function of stress is determined by 

	\begin{equation}
		\label{eq:eq_2}
		\epsilon(t) = \int_{0}^{t} 	J(t- \tau) \frac{d \sigma(\tau)}{d \tau} \hspace{2mm}d\tau,
	\end{equation}

\noindent where $G(t-\tau)$ and $J(t-\tau)$ are the relaxation and creep modulus of the material, respectively. In the frequency domain, the relaxation modulus can be expressed as a complex modulus \(G^*\) at a specified frequency $\omega$, given by: 

	\begin{equation}
		\label{eq:eq_3}
		G^*(\omega) = G'(\omega) + iG''(\omega),
	\end{equation}

\noindent where $G'$ is the storage modulus, a measure of the material's stiffness, and $G''$ is the loss modulus, related to the viscous dissipation of energy due to deformation. The complex modulus, when described by a power-law, with constants $\kappa$ and \(\beta\), is given by

	\begin{equation}
		\label{eq:eq_4}
		G(\omega) =  \kappa (i \omega)^\beta
	\end{equation}

A power-law model of the microstructure of brain white matter (BWM), comprising axons and an ECM, is obtained by fitting the model to experimentally measured shear moduli from oscillatory tests on the porcine optic nerve \cite{ arbogast_fiber-reinforced_1999, WebPlotDigitizer}. 

\begin{equation}
	\label{eq:eq_5}
	\Re(G_{exp}(\omega)) + i\Im(G_{exp}(\omega)) = \kappa (i \omega)^\beta
\end{equation}

The power-law parameters for the axon and ECM are obtained by transforming \cref{eq:eq_5} into the logarithmic  domain and applying linear regression. This is accomplished by defining a cost function in the form, 

\begin{equation}
	\label{eq:eq_6}
	\begin{aligned}
		E &= \frac{1}{2m} \sum_m (\ln(\kappa) + \beta \ln(\omega) - \ln(G'_{exp}(\omega)))^2 \\
		&\qquad + \hspace{1mm} (\beta \frac{\pi}{2} - \ln(G''_{exp}(\omega)))^2, 
	\end{aligned}
\end{equation}

\noindent where \(E\) is the cost function and, m is the number of input frequency points. The cost function is minimized with respect to \(\ln(\kappa)\) and \(\beta\). The log-transformation linearizes the model and makes gradient descent optimization numerically more stable. The cost function is minimized with respect to \(\ln(\kappa)\) and \(\beta\). 

\begin{equation}
	\label{eq:eq_7}
	\begin{aligned}
		&\frac{\partial E}{\partial \ln(\kappa)} = \frac{1}{m} \sum_m (\ln(\kappa) + \beta \ln(\omega) - \ln(G'_{exp}(\omega))) \\
		&\frac{\partial E}{\partial \beta} = \frac{1}{m} \sum_m ((\ln(\kappa) + \beta \ln(\omega) - \ln(G'_{exp}(\omega))) \ln(\omega) \\
		&\qquad + \frac{\pi}{2}(\frac{\beta \pi}{2} -\ln(G''_{exp}(\omega))))
	\end{aligned}
\end{equation}

\noindent The parameters \(\ln(\kappa)\) and \(\beta\) are updated simultaneously during each iteration as shown below. The learning rates \(\gamma_1 \) and \(\gamma_2 \) are adjusted through trial and error. 

\begin{equation}
	\label{eq:eq_8}
	\begin{aligned}
		\ln(\kappa) &:= \ln(\kappa) - \gamma_1 \frac{\partial E}{\partial \ln(\kappa)} \\
		\beta &:= \beta - \gamma_2 \frac{\partial E}{\partial \beta}
	\end{aligned}
\end{equation}

\begin{figure*}[htbp]
	\centering
	\begin{subfigure}[b]{0.48\linewidth}
		\includegraphics[width=\linewidth]{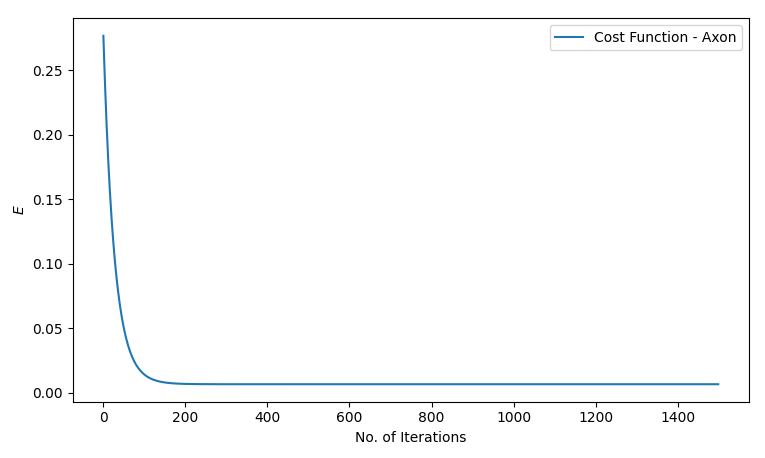}
		\caption{}
	\end{subfigure}
	\hfill
	\begin{subfigure}[b]{0.48\linewidth}
		\includegraphics[width=\linewidth]{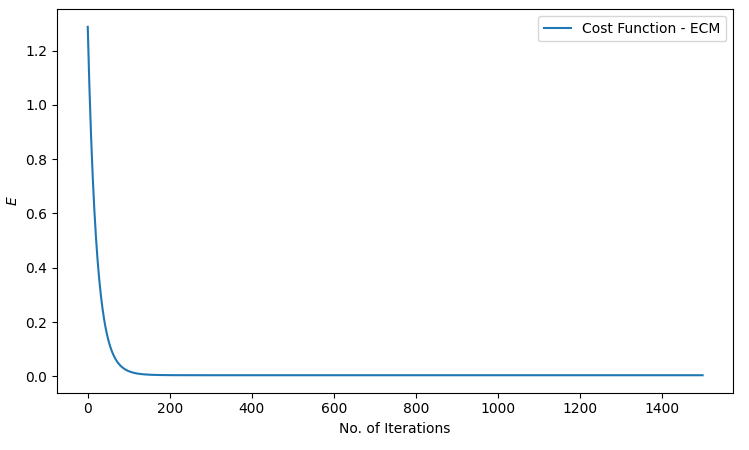}
		\caption{}
	\end{subfigure}
	
	\caption{Evolution of the cost function over iterations for determining the power-law model parameters for (a) axons and (b) ECM.A total of 1500 iterations were performed to ensure convergence and stability of the minimization process.}
	\label{fig:cost_function}
\end{figure*}

\begin{figure*}[h]
	\centering
	\begin{subfigure}[b]{0.48\linewidth}
		\includegraphics[width=\linewidth]{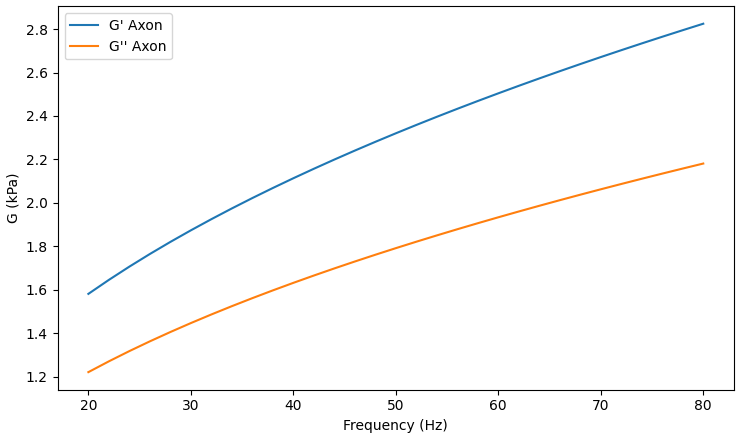}
		\caption{}
	\end{subfigure}
	\hfill
	\begin{subfigure}[b]{0.48\linewidth}
		\includegraphics[width=\linewidth]{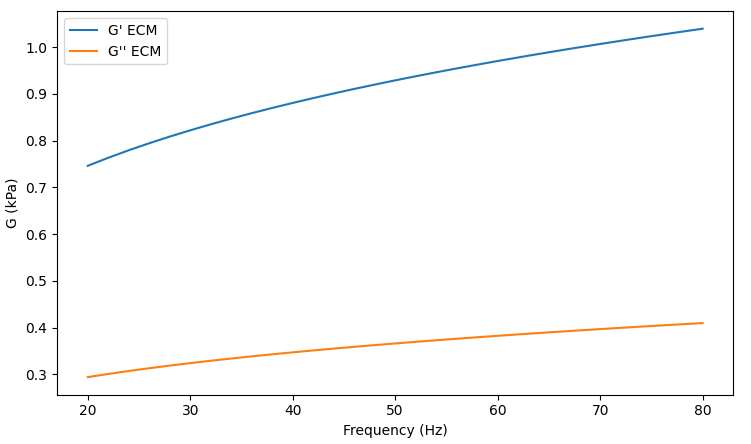}
		\caption{}
	\end{subfigure}
	
	\caption{Storage and loss moduli predicted by the power-law model for (a) axons and (b) ECM as functions of frequency.}
	\label{fig:G_omega}
\end{figure*}

The minimization of the cost function is shown in \ref{fig:cost_function}. The estimated storage and loss moduli for the axons and ECM, obtained from the power-law model are shown in \ref{fig:G_omega}. It can be observed that the axon is stiffer and more viscous than the matrix \cite{arbogast1998material, arbogast_fiber-reinforced_1999}. The power-law parameters for the axon and ECM are listed in \ref{tab:powerlaw}.

\begin{table}[htbp]
	
	\caption{Power-law model parameters for axon and ECM components.}
	\label{tab:powerlaw}
	\centering
	\begin{tabularx}{\columnwidth}{@{\extracolsep{\fill}} lcc}
		\hlineB{2}		
		\textbf{Component} & \boldsymbol{$\kappa (Pa \cdot s^{\beta})$} & {\boldsymbol{$\beta$}} \\		
		\hlineB{2}\hline
		Axon & 264.115 & 0.419 \\
		\hline
		ECM  & 252.503 & 0.239 \\
		\hlineB{2}
	\end{tabularx}
\end{table}

	\subsection{Governing Equations}
	\label{subsec:cmm}
	
	In the time domain, a power-law model with constants $A_\beta$ and $\beta$ can be expressed as, 
	
	\begin{equation}
		\label{eq:eq_9}
		G(t) =  A_\beta t^{-\beta}
	\end{equation}
	
	\noindent Substituting \cref{eq:eq_9} into \cref{eq:eq_1}, we obtain the following, 
	
	\begin{equation}
		\label{eq:eq_10}
		\sigma(t) = \int_{0}^{t} A_\beta (t-\tau)^{-\beta}\frac{d \epsilon(\tau)}{d \tau} \hspace{2mm}d\tau
	\end{equation}

	\Cref{eq:eq_10} is of a form commonly encountered in the field of fractional calculus and is known as the Riemann–Liouville integral. It represents one of the most widely used definitions of the fractional derivative. Consider the function \(u(t)\). Repeated integration of \(u(t)\), performed \(\alpha \in \mathbb{R}\) number of times, can be expressed using Cauchy's integral formula as:
	
	\begin{equation}
		\label{eq:eq_11}
		\begin{aligned}
		I^{\alpha}u(t) &= \int_{0}^{t} \int_{0}^{t} \int_{0}^{t} ... u(t) dt^\alpha \\
		&= \frac{1}{\Gamma(\alpha)} \int_0^t (t -\tau)^{\alpha -1} u(\tau) d{\tau},
		\end{aligned}
	\end{equation} 
	
	\noindent where $I^\alpha\{\cdot\}$ denotes repeated integration over $\alpha$ number of times, $\Gamma(\alpha)$ is the Gamma function. If the RHS of \cref{eq:eq_10} is set to $f(\tau)$, where \(f(0) = 0 \), then differentiating $f(\tau)$, $\alpha$ number of times yields the original function, $u(\tau)$.
	
	\begin{equation}
		\label{eq:eq_12}
		\begin{aligned}
			f(\tau) &=\frac{1}{\Gamma(\alpha)} \int_0^t (t -\tau)^{\alpha -1} u(\tau) d{\tau}, \\ \\
			u(\tau) &= D^\alpha f(\tau),
		\end{aligned}		
	\end{equation}

\noindent where $D^\alpha\{\cdot\}$ denotes differentiation by order $\alpha$, a non-integer value. Performing a Laplace transform of \cref{eq:eq_12} yields \(\frac{U(s)}{s^\alpha} = F(s)\). Rewriting the expression as \(U(s) = s(\frac{F(s)}{s^{(1-\alpha)}})\) and then taking the inverse Laplace transform followed by integrating by parts, we obtain:

\begin{equation}
	\label{eq:eq_13}
	D^\alpha f(t) = \frac{1}{\Gamma(1-\alpha)} \int_0^t (t - \tau)^{-\alpha} \frac{d f(\tau)}{d \tau}
\end{equation}

\noindent Comparing \cref{eq:eq_10} and \cref{eq:eq_13}, the viscoelastic stress can be written as,

\begin{equation}
	\label{eq:eq_14}
	\sigma(t) = C_\beta D^\beta \epsilon(t)
\end{equation}

\noindent where $C_\beta = A_\beta \Gamma(1-\alpha)$.Taking the Laplace transform of \cref{eq:eq_10} yields $G(s) = \frac{s A_\beta}{s^{1 - \beta}}$, where $G(s)$ is the shear modulus in the Laplace domain. Considering the analytical form of the Laplace and Fourier transforms,

\begin{equation}
	\label{eq:eq_15}
	\begin{aligned}
		\mathcal{L} [G(t)] &= G(s) = \int_{0}^{\infty}  G(t) e^{-st} \hspace{2mm}dt,\\
		\vspace{2mm}
		\mathcal{F} [G(t)] &= G(\omega) = \int_{-\infty}^{\infty}  G(t) e^{-i \omega t} \hspace{2mm}dt,
	\end{aligned}
\end{equation}

\noindent and given $G(t)$ is causal, i.e. \(G(t) = 0 \hspace{2mm} \forall \hspace{2mm} t<0\), $\mathcal{L} [G(t)]$ exists with a region of convergence that includes the imaginary axis, the Laplace and Fourier transform are related via the imaginary unit \(i = \sqrt{-1}\). Therefore, \(s\) can be interchanged with \(i \omega\), where \(\omega\) is the frequency in rad/s. This implies, from  \cref{eq:eq_4} and \cref{eq:eq_15}, that $C_\beta = \kappa$.

\subsection{Fractional Viscoelastic Constitutive Material Model}

Given that the stress is proportional to the zeroth order derivative of strain in an elastic spring and to its first order derivative in a dashpot, G.W. Scott Blair posited an intermediate material---between a Hookean solid and a Newtonian fluid---in which the stress is proportional to non-integer order of strain \cite{caffyn_significance_1945, scott_blair_role_1947, noauthor_survey_1999}. The mechanical response of such a \enquote{spring-pot} can be described by employing the Riemann-Liouville integral in \cref{eq:eq_10} and, \cref{eq:eq_14} as \cite{fedorchenko_introduction_2008, bagley_theoretical_1983, bonfanti_fractional_2020}:

	\begin{equation}
		\label{eq:eq_16}
		\begin{aligned}
			\sigma(t) &= \kappa \frac{d^\beta \epsilon(t)}{dt^\beta} \\
			&= \frac{\kappa}{\Gamma(1-\beta)} \int_{0}^{t}  (t-\tau)^{-\beta}\frac{d \epsilon(\tau)}{d \tau} \\
			& \hspace{30mm}  \forall \hspace{1mm} (0 \leq \beta \leq 1)
		\end{aligned}
	\end{equation}
	
	The fractional spring-pot model for the microstructure of BWM is implemented in three dimensions using the finite element method in Abaqus software. In 3-dimensions, the relaxation matrix can be split into its volumetric and deviatoric components \cite{alotta2017behavior},
	
	\begin{equation}
		\label{eq:eq_17}
		\begin{aligned}
		G_{ijkm}(t) &= (K_R(t) - \frac{2}{3} G_{R}(t)) \delta_{ij} \delta_{km} \\
		&+ G_{R}(t)(\delta_{ik} \delta_{jm} + \delta_{im} \delta_{jk}),
		\end{aligned}
	\end{equation}

\noindent where $\delta_{ij}$ is the Kronecker delta function. $G_R(t)$ and $K_R(t)$ are the deviatoric and volumetric power-law functions respectively. Substituting \cref{eq:eq_17} into \cref{eq:eq_16} and rewriting terms of the power-law coefficient $\kappa$, in terms of the volumetric and deviatoric components, $K_\beta$ and $G_\beta$, the stress-strain equation becomes,

\begin{equation}
	\label{eq:eq_18}
	\begin{aligned}
	\sigma_{ij} &= \frac{1}{\Gamma(1-\beta)}\int_{0}^{t} (K_\beta - \frac{2}{3} G_\beta) \delta_{ij}  \dot{\epsilon_{kk}}(t) dt \\
	&+ \frac{1}{\Gamma(1-\beta)}\int_{0}^{t} G_\beta (\dot{\epsilon_{ij}}(t) + \dot{\epsilon_{ji}}(t)) dt
	\end{aligned}
\end{equation}

For any function $f(x)$ containing $\alpha$ order continuous derivatives, the fractional derivative of $f(x)$ can be computed by the Grunwald-Letnikov (GL) operator which is given by \cite{199941, schmidt_no_2002},

\begin{equation}
	\label{eq:eq_19}
	\begin{aligned}
		\frac{d^\alpha f(x)}{dx^\alpha} &= \lim_{n\to\infty} (\frac{n}{x})^\alpha \sum_{l=0}^{n} (-1)^l \frac{\Gamma(\alpha+1)}{\Gamma(\alpha+1-l)} f(x-\frac{l}{n}x)
	\end{aligned}
\end{equation}

Under weak conditions imposed on $f(x)$ as described above, the Riemann-Liouville integral is equivalent to the GL definition of the fractional derivative. Alotta et al. \cite{alotta2018finite} implemented a 3D fractional viscoelastic user material (\texttt{UMAT}) subroutine in Abaqus using the GL operator. For a 3D state of stress, $\sigma\left(t\right)={[\sigma}_{11}\ \sigma_{22\ }\sigma_{33}\ \tau_{23}\ \tau_{31}\ \tau_{12}]$, the stress in any direction for \cref{eq:eq_18} can be obtained by its GL equivalent as,

\begin{equation}
	\label{eq:eq_20}
	\begin{aligned}
	\sigma_{ij}^{k+1}&={(K}_\beta\ -\ \frac{2G_\beta}{3})\left(\frac{1}{\Delta t}\right)^\beta\sum_{l=1}^{k+1}\varphi_l\epsilon_{kk}\left(\left(k+1-l\right)\Delta t\right)\\
	&+2G_\beta\left(\frac{1}{\Delta t}\right)^\beta\sum_{l=1}^{k+1}\varphi_l\epsilon_{ij}((k+1\ -\ l)\Delta t),
	\end{aligned}
\end{equation}

\noindent where $k = \frac{Total \hspace{1mm} time}{\Delta t}$ is the number of iterations, $\epsilon_{kk}$ is the volumetric strain, and $\phi_l$ are the Grunwald coefficients used to handle the gamma functions in \cref{eq:eq_19}, which can be computed as described in \cite{1999199} as,

\begin{equation}
	\label{eq:eq_21}
	\varphi_{l+1} = \frac{(k - 1 -\beta)}{k} \varphi_l,	 \hspace{5mm} \varphi_1 = 1.
\end{equation}

\Cref{eq:eq_20} is implemented in Abaqus through a vectorized Fortran user material subroutine \texttt{VUMAT}, which is specifically used for explicit time integration.  Note that the calculation of stress for any given increment requires storing and accessing the history of strains for all previous increments. This makes the simulation of large models using an explicit integration scheme computationally expensive. The \texttt{VUMAT} is invoked multiple times by the solver during each increment as it processes blocks of material points. For each material point, the strain history must be indexed, stored, and retrieved based on the corresponding element number and integration point. 
\subsection{Finite Element Model}

\begin{figure}[!h]
	\centering
	\includegraphics[width=0.48\textwidth]{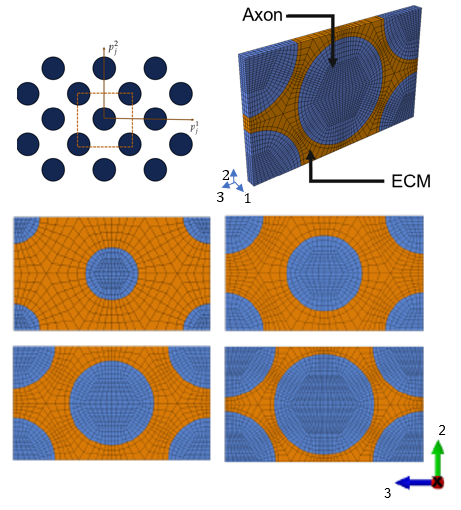}
	\caption{\textsl{Left, Top:}A schematic of a periodic geometry of axons in an ECM. \textsl{Right, Top:} A FE model of the hexagonally packed RVE of two axons embedded in an ECM, \textsl{Bottom:} RVEs with different volume fractions [Clockwise: 0.2, 0.3, 0.5, 0.7].}\label{fig:image5}
	\end{figure}

The FE model of the microstructure of BWM is constructed by assuming a periodic distribution of hexagonally packed axons embedded in an ECM as shown in \ref{fig:image5}. The diameter of the axon in the RVE is 10 $\mu$m. The volume fraction is varied in steps of $0.05$ from $0.2$ to $0.7$

Periodic boundary conditions (PBC) are applied to the hexagonally packed structure. PBC is imposed on the nodes of the model through a set of linear constraint equations in Abaqus \cite{tian2019periodic, sadaba_special-purpose_2019}.
The far-field gradient is applied through the degrees of freedom of reference points (RP) using the following equation \cite{abq}, 

\begin{equation}
	\label{eq:eq_22}
	u_i(x_j + p_j^\alpha) = u_i(x_j) + \frac{\partial u_i}{\partial x_j} p_j^\alpha,
\end{equation}

\noindent where $x_j$ is a nodal coordinate,  $ p_j^\alpha$ is the $\alpha$th vector of periodicity, and $\frac{\partial u_i}{\partial x_j}$ is the far-field gradient of the displacement. RP nodes are not physically connected to any elements in the model. However, the equivalent internal force response in a given direction can be obtained by extracting the reaction forces at a prescribed RP. To correctly apply PBC, it is essential to ensure that the mesh on each pair of opposite faces is identical. This allows the desired field variable at $x_j$ to be paired with its counterpart $x_j + p_j^\alpha$ on the opposing face when formulating the linear constraint equations \cite{abq, omairey2019development} shown below

\begin{equation}
	\label{eq:eq_23}
	A_1 u^p_i + A_2 u^q_j + \dots + A_n u^v_s = 0, 
\end{equation}

\noindent where $n$ is the number of terms in the equation, $u^P_i$ is the displacement variable at node $P$ and degree of freedom $i$, and $A_n$ are the coefficients.

The finite element model of the RVE, incorporating  the fractional viscoelastic \texttt{VUMAT}, is subjected to a quasi-static displacement boundary condition along six loading directions with a maximum strain of $0.5 \%$. An explicit time integration scheme is used to solve the model. The homogenized stress obtained from each loading case is then passed to an optimization workflow. 

	\subsection{Model Validation}
	\label{subsec:mv}
	
\begin{figure}[ht]
	\centering
	\begin{subfigure}[b]{0.48\columnwidth}
		\includegraphics[width=\linewidth]{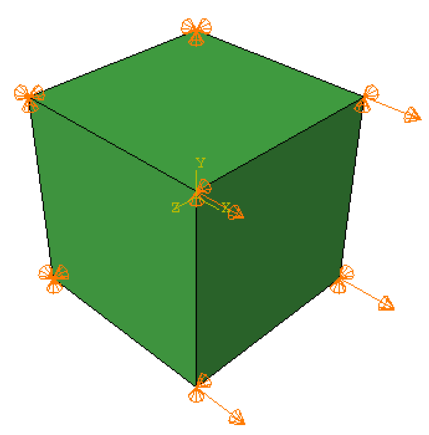}
		\caption{}
		\label{subfig:uni_stretch}
	\end{subfigure}
	\hfill
	\begin{subfigure}[b]{0.48\columnwidth}
	\includegraphics[width=\linewidth]{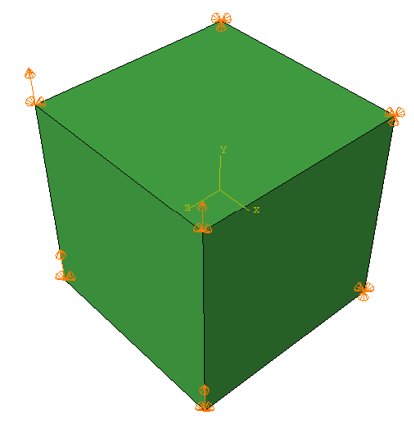}
	\caption{}
	\label{subfig:cube_shear}
	\end{subfigure}
	\hfill
	\begin{subfigure}[b]{\columnwidth}
		\includegraphics[width=\linewidth]{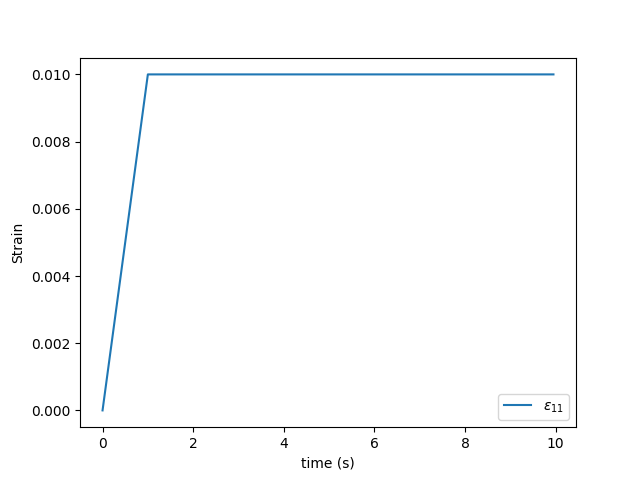}
		\caption{}
		\label{subfig:disp}
	\end{subfigure}
	\caption{
		Benchmark testing of the \texttt{VUMAT} subroutine using a unit cube model. 
		(a) Unit cube subjected to a quasi-static tensile stretch as described in (c).
		(b) Unit cube subjected to a quasi-static shear deformation as described in (c).
		(c) Quasi-static strain loading profile applied to (a) and (b), with an amplitude of $1 \% $.
	}
	\label{fig:unit_cube}
\end{figure}

	The \texttt{VUMAT} subroutine is tested and validated using a set of benchmark problems with known analytical solutions. We first apply the model to a single element unit cube under uniaxial stretch (\Cref{{subfig:uni_stretch}}) and shear (\Cref{subfig:cube_shear}). The material properties for the cube are based on test case developed by Alota et al \cite{alotta2018finite, hesammokri2019implementation}. For a quasi-static displacement boundary condition, such as a uniaxial stretch, the strain can be written in terms of a unit-step function as, 
	
	\begin{equation}
		\label{eq:eq_24}
		\begin{aligned}
			\epsilon_{11} &= \bar{\epsilon} \left[ t \mathcal{H}(t) - (t-1) \mathcal{H}(t-1)\right]\\
			\epsilon_{22} &= \epsilon_{33} = 0
		\end{aligned}		
	\end{equation}

\noindent where $\bar{\epsilon}$ is a scalar and $\mathcal{H}(t)$ is the Heaviside function. In this case, the quasi-static boundary condition is applied such that the strain increases linearly with time until $t = 1s$, after which it remains constant (see \Cref{subfig:disp}). The time-derivative of strain is a rectangular pulse in the interval $t \in [0,1]s$.  By substituting \cref{eq:eq_24} in \cref{eq:eq_18}, the stress along the direction of the stretch is obtained as, 

\begin{equation}
	\label{eq:eq_25}
	\begin{aligned}
	\sigma_{11} &= \frac{\bar{\epsilon}}{\Gamma(1-\beta)} \left(K_\beta + \frac{4G_\beta}{3}\right)\\
	&\left[ t^{(1-\beta)}\mathcal{H}(t) - (t-1)^{(1-\beta)}\mathcal{H}(t-1)\right]
	\end{aligned} 
\end{equation}
 
\begin{figure}[h!]
	\centering
	\includegraphics[width=\columnwidth]{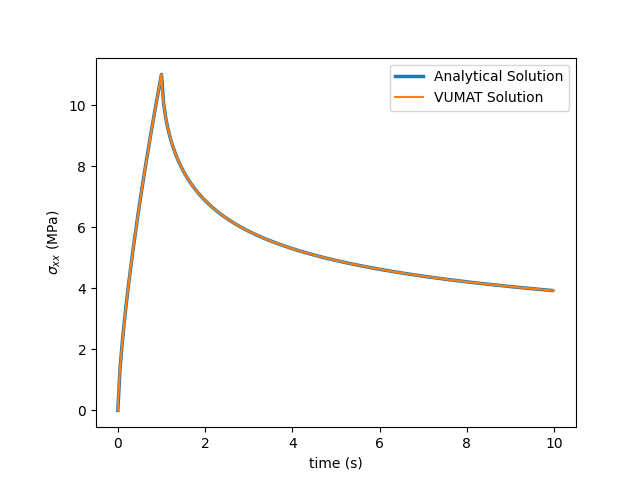}
	\caption{Comparison of the uniaxial stress response between the analytical solution and \texttt{VUMAT} obtained under a quasi-static strain with a peak amplitude of 0.01.}\label{fig:cube_tensile_stretch}
	\label{fig:uniaxial_stress}
\end{figure}

\begin{figure}[h!]
	\centering
	\includegraphics[width=\columnwidth]{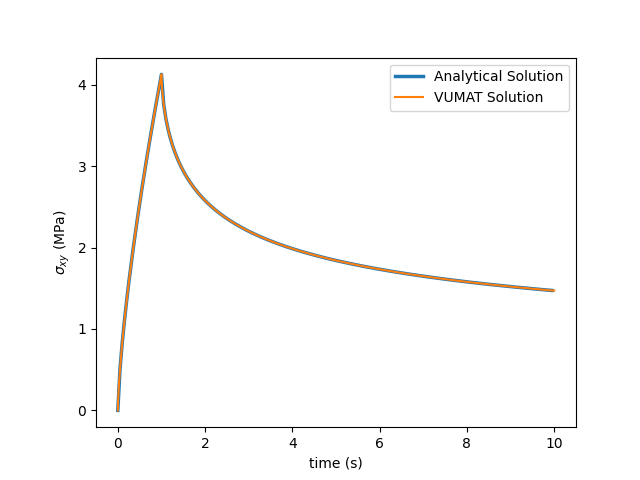}
	\caption{Comparison of the shear stress response between the analytical solution and \texttt{VUMAT} obtained under a quasi-static strain with a peak amplitude of 0.01.}
	\label{fig:cube_shear_stretch}
\end{figure}

\begin{figure*}[h]
	\centering
	\begin{subfigure}[b]{0.3\linewidth}
		\includegraphics[width=\linewidth]{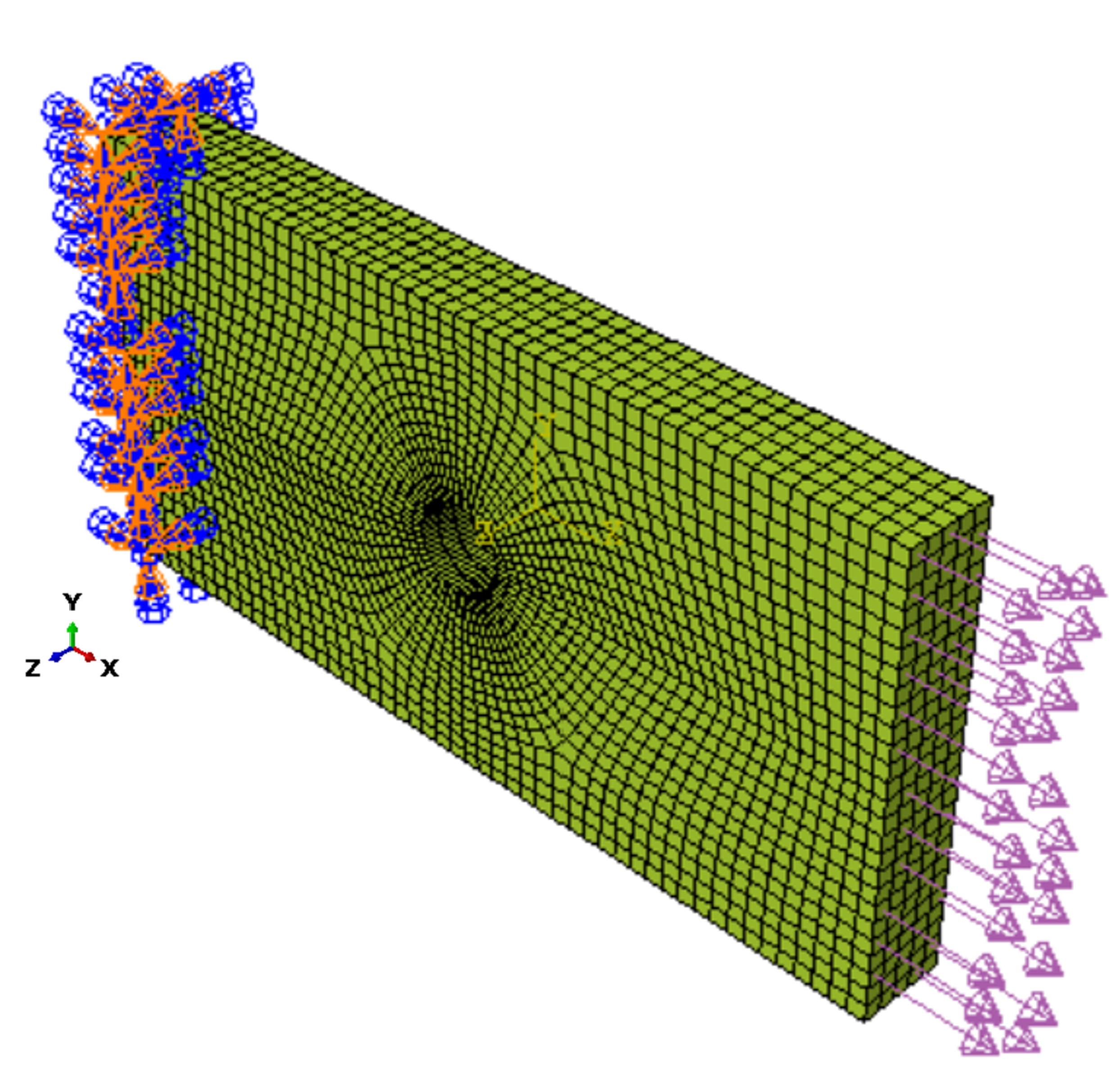}
		\caption{}
		\label{subfig:plate_model}
	\end{subfigure}
	\hfill
	\begin{subfigure}[b]{0.3\linewidth}
		\includegraphics[width=\linewidth]{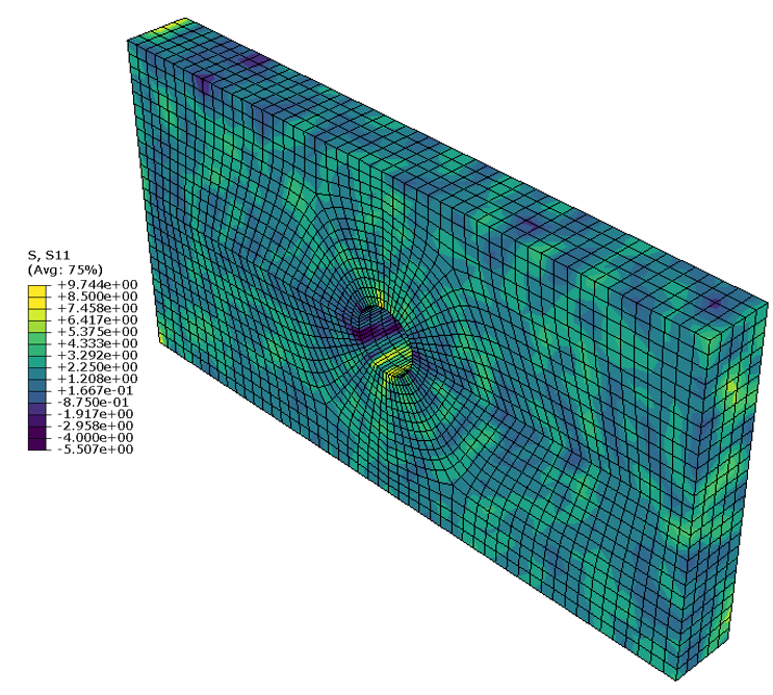}
		\caption{}
		\label{subfig:Plate_with_hole_model_stress}
	\end{subfigure}
	\hfill
	\begin{subfigure}[b]{0.35\linewidth}
		\includegraphics[width=\linewidth]{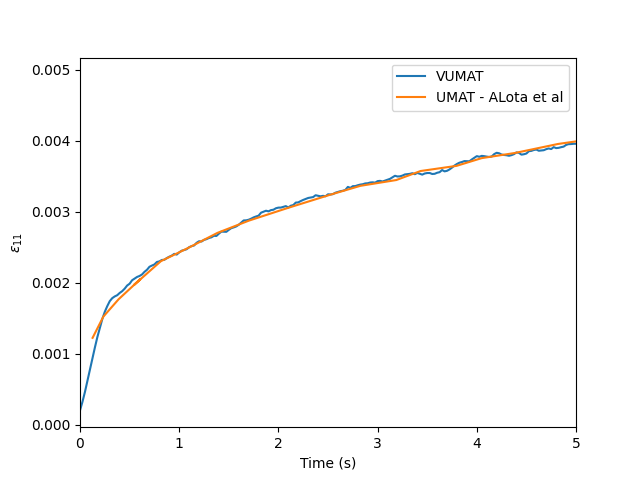}
		\caption{}
		\label{subfig:plate_strain_plot}
	\end{subfigure}
	\caption{
		Benchmark testing of the \texttt{VUMAT} subroutine using a plate-with-a-hole model under uniaxial tension. 
		(a) Finite element model constrained on the left end with a uniform tensile load of 2MPa applied on the right. The model contains 8720 C3D8R elements (Hexahedral elements with reduced integration.)
		(b) Fringe plot of the stress distribution, \(\sigma_{11}\) at \(t=5s\).
		(c) Comparison of the evolution of tensile strain with time in the solid model between the \texttt{VUMAT} solution implemented in the current study and the \texttt{UMAT} subroutine developed by Alota et al \cite{alotta2018finite}.
	}
	\label{fig:plate_with_hole_model}
\end{figure*}

The comparison between the analytical solution and the \texttt{VUMAT} subroutine is shown in \ref{fig:uniaxial_stress}. The shear stress response under a pure-shear type displacement is shown in \ref{fig:cube_shear_stretch}. Next, a plate with a hole is subjected to a uniform tensile load of 2 MPa held constant for $t=5$s. The evolution of the tensile strain in the plate computed via the \texttt{VUMAT} implementation is compared with the \texttt{UMAT} subroutine developed by Alota et al \cite{alotta2018finite} (see \Cref{fig:plate_with_hole_model}).  

\begin{figure}[h!]
	\centering
	\begin{subfigure}[b]{\columnwidth}
		\includegraphics[width=\linewidth]{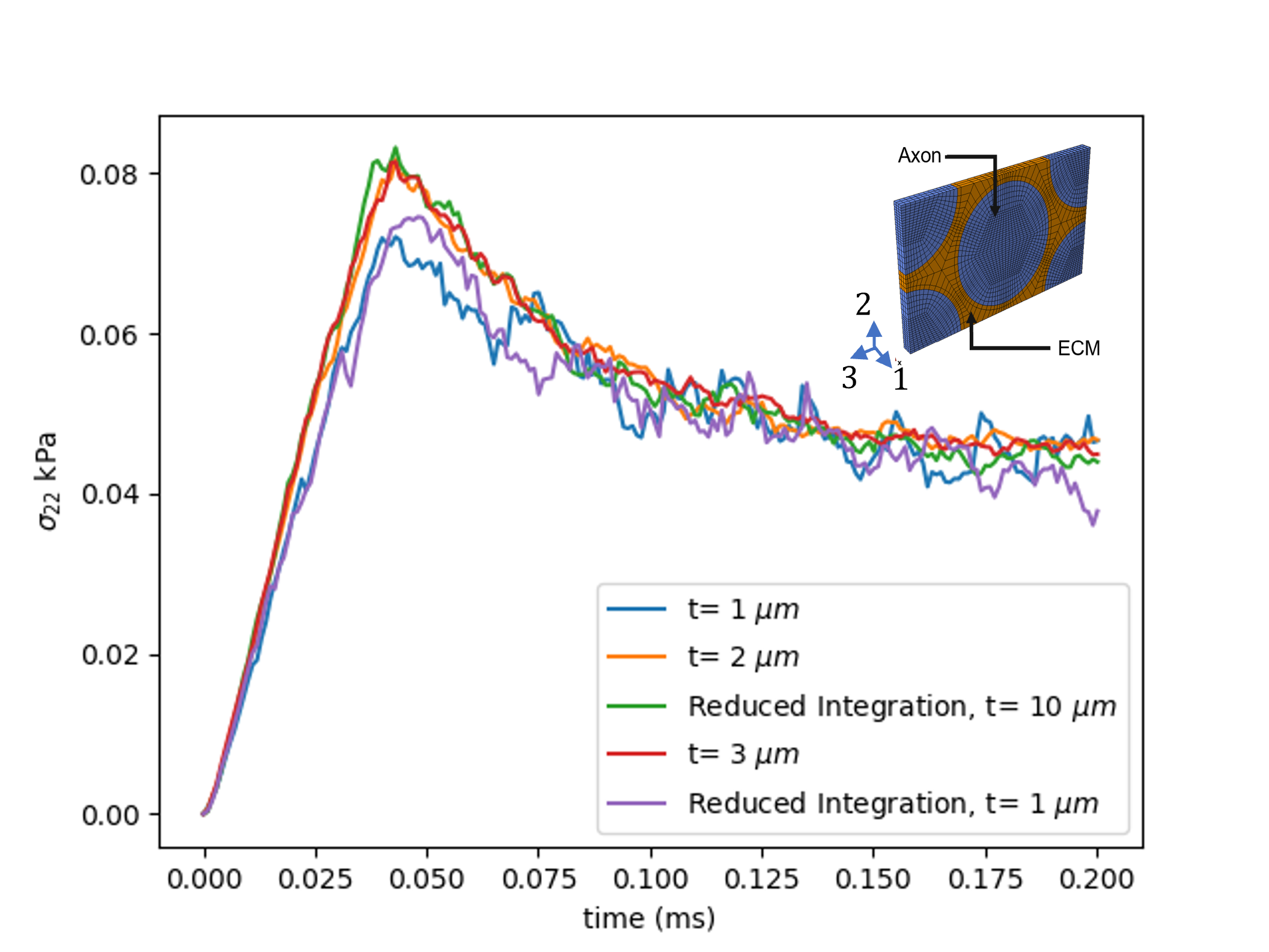}
		\caption{}
	\end{subfigure}
	\begin{subfigure}[b]{\columnwidth}
		\includegraphics[width=\linewidth]{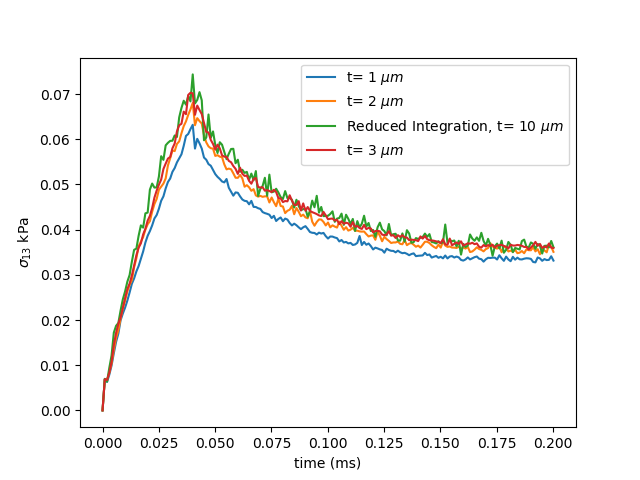}
		\caption{}
	\end{subfigure}
	\caption{(a) Comparison of stress along, (a) transverse normal direction 2-2 ($\sigma_{22}$) (b) shear along direction 1-3 ($\sigma_{13}$), under a quasi-static deformation with peak strain of $0.5 \% $ for different RVE thickness' and number of elements, applied to an RVE with $vf=0.4$}.
	\label{fig:rve_depth_stress}
\end{figure}

Typically, the size of the RVE must be large enough to adequately represent the macroscale behavior. However, in practice, the size, specifically the thickness of the RVE along the fiber direction is not always obvious. In a hexagonally packed structure, given the volume fraction and the fiber radius, the cross-sectional area of the RVE becomes fixed. When periodic boundary conditions are applied, the strain field is uniform along the fiber direction and is independent of the thickness of RVE. However, the same cannot be said for the other loading directions. While RVE's with insufficient depth may fail to capture the mechanical behavior, excessively thick RVEs significantly increase computational cost without necessarily providing additional insight. In order to determine the optimal size of the RVE, we consider different depths and number of elements/integration points shown in \ref{tab:rve_depth} for a volume fraction, $vf=0.4$. The stress response under elongation in 2-2 and shear along 1-3 directions are shown in \ref{fig:rve_depth_stress}. It can be observed that the RVE with a thickness of $1 \mu m$ using both full and reduced integration, underpredicts the stress along the 2-2 direction. However, the $2 \mu m$, $3 \mu m$ and $10 \mu m$ model with reduced integration, all converge toward a common solution. Similarly for shearing along 1-3, the $1 \mu m$ model underpredicts the magnitude of the shear stress but the $2 \mu m$, $3 \mu m$ and $10 \mu m$ model with reduced integration converge toward a common stress solution indicating the $2 \mu m$ thickness model is optimal. 

\begin{table}[h]
	
	\caption{Summary of RVE depths and element types tested for determining the optimal RVE size.}
	\label{tab:rve_depth}
	\centering
	\begin{tabularx}{\columnwidth}{@{\extracolsep{\fill}} lccc}
		\hlineB{2}		
		\textbf{RVE Thickness} & \textbf{Element} & \textbf{Number of} \\		
		\boldsymbol{$(\mu m)$} & \textbf{Type} & \textbf{Elements}\\
		\hlineB{2}\hline
		1 & C3D8 & 1952 \\
		\hline
		1  & C3D8R & 1952 \\
		\hline
		2 & C3D8 & 3904 \\
		\hline
		3  & C3D8 & 5856 \\
		\hline
		10  & C3D8R & 9760 \\
		\hlineB{2}
\	\end{tabularx}
\end{table}

\begin{figure}[h!]
 	\centering
 	\begin{subfigure}[b]{0.49\columnwidth}
 		\includegraphics[width=\linewidth]{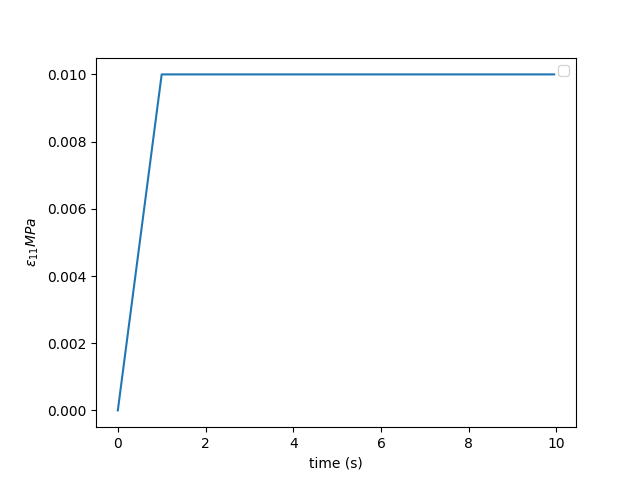}
 		\caption{}
 		\label{subfig:stretch}
 	\end{subfigure}
 	\begin{subfigure}[b]{0.49\columnwidth}
 		\includegraphics[width=\linewidth]{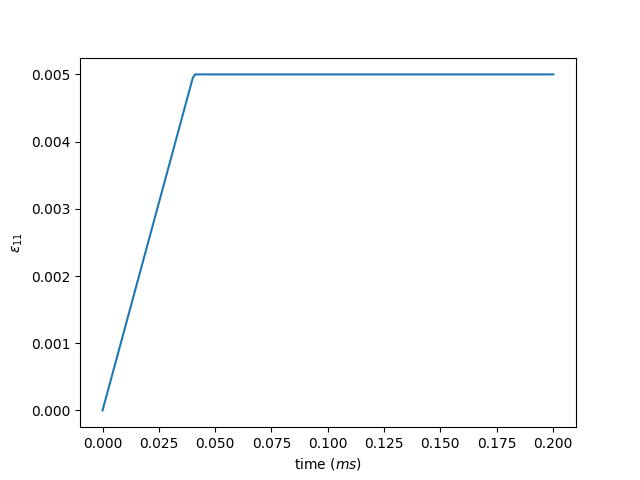}
 		\caption{}
 		\label{subfig:epsilon_11}
 	\end{subfigure}
	 \caption{(a) Quasi-static strain with a maximum magnitude of $1\%$ applied to the unit cube model. (b) Quasi-static strain with a maximum magnitude of $0.5\%$ applied to an RVE with $vf=0.4$. The strain is increased at a rate of 125 $\frac{1}{s}$ for 0.04 $ms$ and kept constant thereafter.}
 \label{fig:rmem-disp}
\end{figure}

\begin{figure}[h!]
 	\begin{subfigure}[b]{\columnwidth}
 		\includegraphics[width=\linewidth]{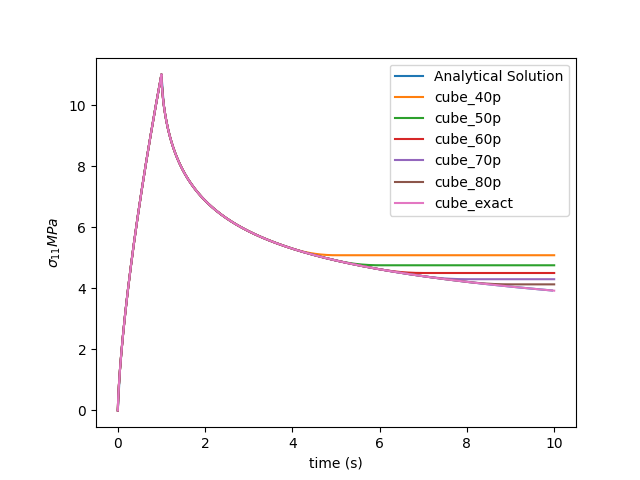}
 		\caption{A comparison of the tensile stress response of a unit cube between the exact solution and simulations with varying fractions of the total time as memory length, $L$.}
 		\label{subfig:Cube_rmem}
 	\end{subfigure}
 	\hfill
 	\begin{subfigure}[b]{\columnwidth}
 		\includegraphics[width=\linewidth]{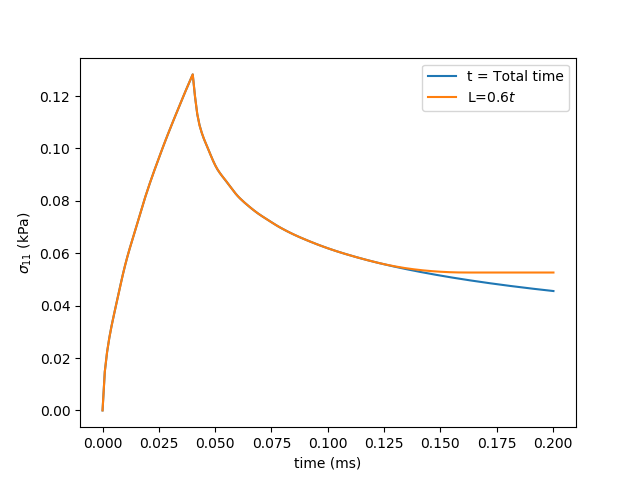}
 		\caption{A comparison of the tensile stress response along the fiber direction for an RVE with $vf = 0.4$, considering the full time history and a reduced-memory simulation with a memory length, L set to $60\%$ of the total simulation time.}
 		\label{subfig:RVE_rmem}
 	\end{subfigure} 
 \caption{An implementation of the \enquote{short-memory} principle. For the quasi-static deformation shown in \Cref{fig:rmem-disp}, all tested values of the memory length L reproduce the maximum stress in the model. Deviation from the exact solution occurs during the relaxation phase of the stress for larger values of L. Similarly for the RVE with BWM properties and vf=0.4, the deviation between the reduced-memory and full time-history simulations is $2\%$. }
 \label{fig:rmem}
\end{figure}

 A key drawback of the fractional viscoelastic model is the need to store and retrieve the entire strain history in order to calculate the stress at the end of each increment. As a result, in an explicit time integration scheme, which typically requires small time increments, the number of computations can become prohibitively large. For large $t$, the role of the history of the strains near the lower terminal, i.e., near $t=0$ can be neglected under certain assumptions. Podlubny \cite{1999199} describes a method known as the \enquote{short-memory} principle which can be used to truncate the memory of strains required to compute the stress in \cref{eq:eq_20}. Alotta et al. \cite{alotta2018finite} also discuss examples of truncated memory for explicit simulations. While this reduces the cost of computation, it also comes at the price of decreased accuracy. The optimal memory length must be determined for each load profile based on the desired level of accuracy. We compute the reduced memory solution by essentially replacing the increasing number of terms in the summation in \cref{eq:eq_20} with a fixed value as the simulation progresses as shown below, 
 
 \begin{subequations}
 	\begin{alignat}{2}
 		\qquad \frac{d^\beta f}{dt^\beta} =& \left(\frac{1}{\Delta t}\right)^\beta\sum_{l=1}^{M+1}\varphi_lf\left(\left(k+1-l\right)\Delta t\right)\label{eq:optProb}\\
 		\text{where } M =& \text{min} \{k, \frac{L}{\Delta t}\}, \hspace{2mm} L = \textsl{Memory Length}
 	\end{alignat}
 \end{subequations} 
 
 \noindent where $M$ is the minimum value between $k=\frac{Total \hspace{1mm} time}{\Delta t}$ and  $\frac{L}{\Delta t}$. Therefore, the number of terms in the summation during the early stages of the simulation is set to $k$ but as the simulation time grows, the terms in the summation is switched to  \enquote{M} which is a fixed number of increments. The application of short-memory principle is demonstrated in \ref{fig:rmem} using a unit cube model with varying fractions of the total simulation time as the memory length and, an RVE with $vf=0.4$ and a memory length corresponding to $60 \%$ of the total time both, under quasi-static boundary conditions. For a quasi-static deformation (see \ref{fig:rmem-disp}), all tested values of the memory length, L reproduce the maximum stress in the model. Deviation from the exact solution occurs during the relaxation phase of the stress for larger values of L. 
 
 For an RVE with $vf=0.4$, containing 3904 hexahedral elements (C3D8) with full integration (8 integration points), a memory length of $60 \%$ of the total simulation time results in a deviation of $2.3\%$ between the reduced-memory and full time-history simulations. The computing time on a single processor with 40 GB of RAM is reduced from 52 hours to 46 hours. 
 
 An optimization-based calibration workflow is used to determine the homogenized fractional viscoelastic parameters for RVEs with varying volume fractions and along each loading direction. Nelder-Mead downhill simplex algorithm is used to determine the material parameters. Simplex is a geometric figure in $ \mathbb{R}^n$ with $n+1$ vertices in an n-dimensional space. It compares the values of the objective function at $N+1$ vertices and gradually moves the polyhedron towards the optimal point by iteratively replacing the worst vertex with a point moved through the centroid of the remaining N points \cite{andersson2012derivative}. 
 
 The workflow implemented using modeFRONTIER \cite{mf}, computes the fractional viscoelastic stress for a set of input spring-pot parameters and minimizes the root mean square deviation (RMSD) between the computed stress and the homogenized stress from the FEM simulation. The optimization problem is formulated as shown in \cref{eq:eq_27}, where $J$ is the cost function, $\sigma_{sim}$ is the FEM solution and $\sigma_{c}$ is the computed stress. 
 
 \begin{subequations}
 	\label{eq:eq_27}
 	\begin{alignat}{2}
 		&\!\min        &\qquad& J = \sqrt{\frac{1}{m}\sum_{n=1}^m \frac{(\sigma_{sim} - \sigma_{c})^2}{\sigma_{sim}^2}} \label{eq:optProb}\\
 		&\text{subject to} &      & 0 \leq \beta \leq 1 ; \hspace{3mm}m > 0 \label{eq:constraint1}
 	\end{alignat}
 \end{subequations}
 
	\section{Results}
	\label{sec:Res}

\begin{figure*}[h!]
	\centering
	
	\begin{subfigure}[b]{0.3\linewidth}
		\includegraphics[width=\linewidth]{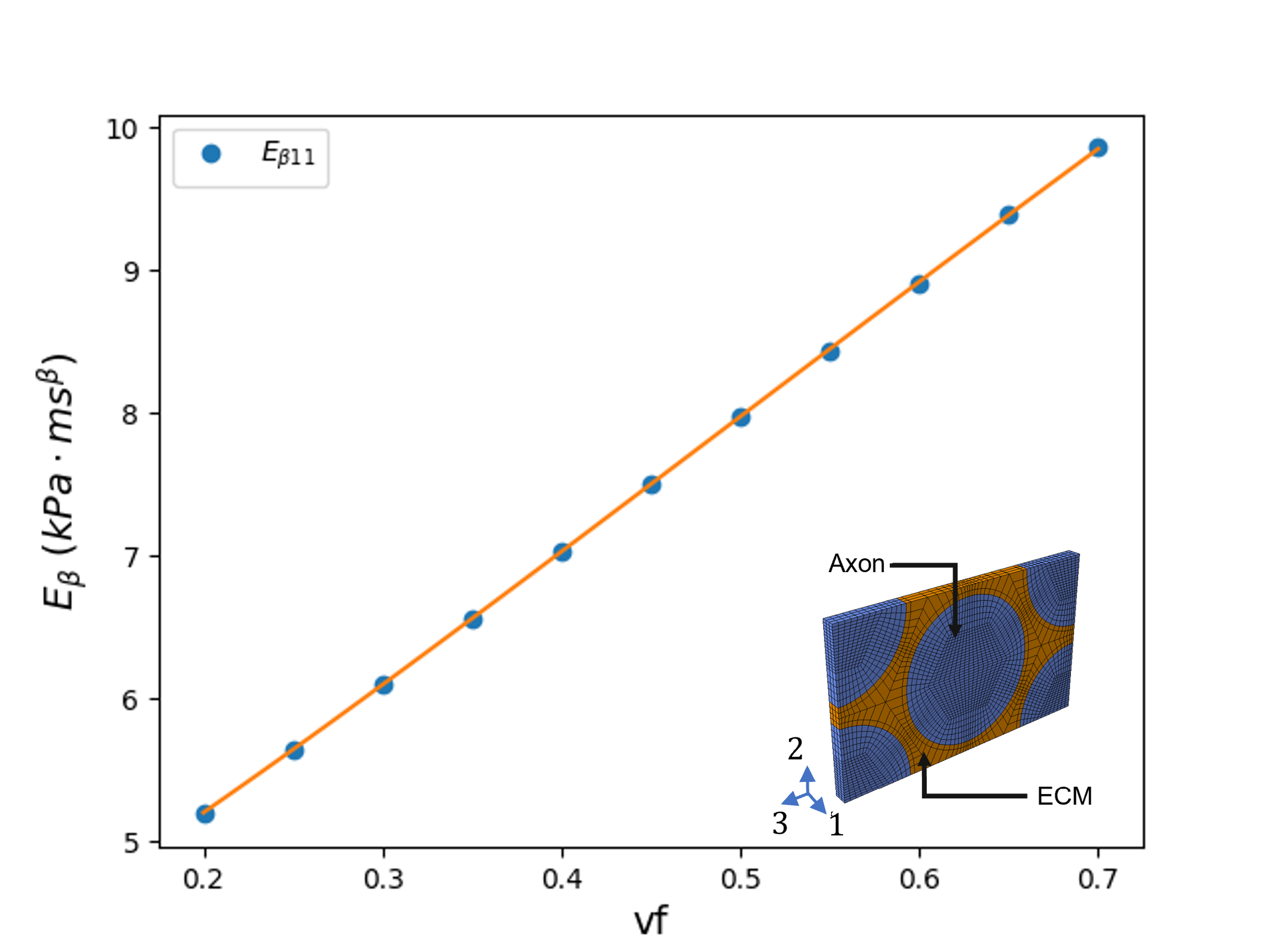}
		\caption{}
	\end{subfigure}
	\hfill
	\begin{subfigure}[b]{0.3\linewidth}
		\includegraphics[width=\linewidth]{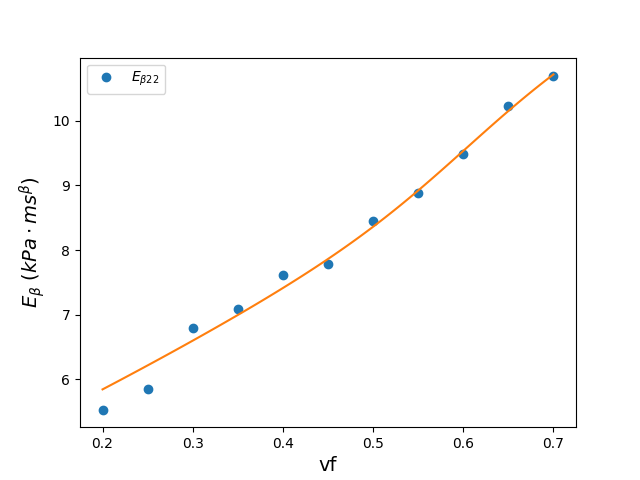}
		\caption{}
	\end{subfigure}
	\hfill
	\begin{subfigure}[b]{0.3\linewidth}
		\includegraphics[width=\linewidth]{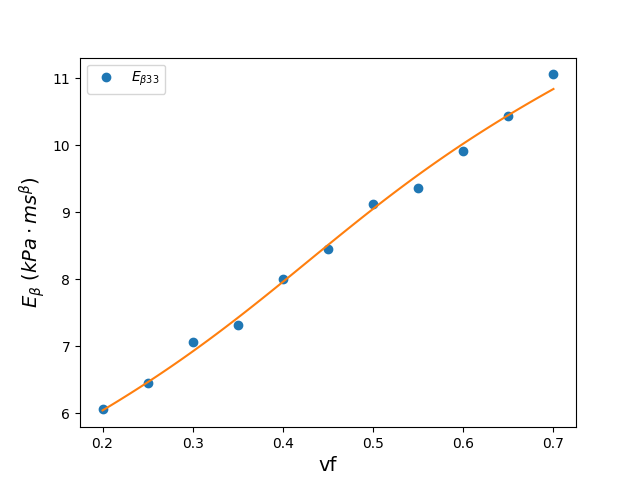}
		\caption{}
	\end{subfigure}
	
	\vspace{0.4cm}
	
	\begin{subfigure}[b]{0.3\linewidth}
		\includegraphics[width=\linewidth]{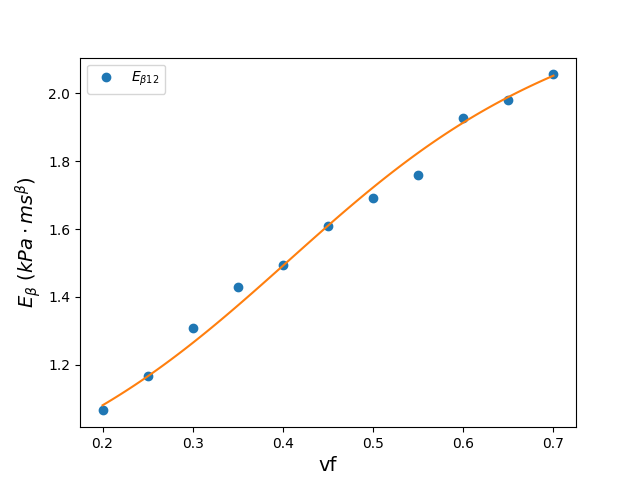}
		\caption{}
	\end{subfigure}
	\hfill
	\begin{subfigure}[b]{0.3\linewidth}
		\includegraphics[width=\linewidth]{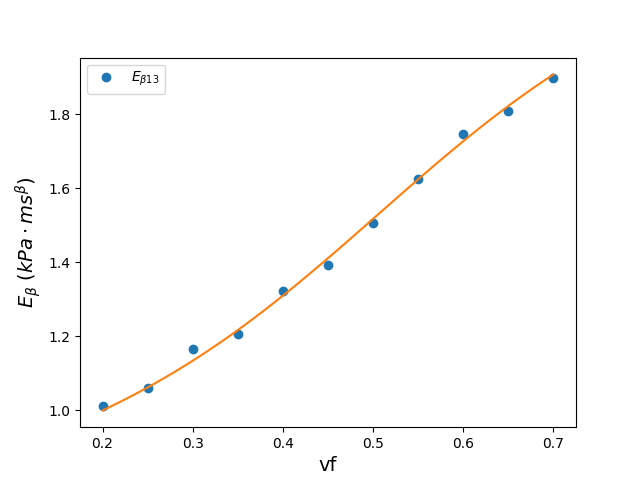}
		\caption{}
	\end{subfigure}
	\hfill
	\begin{subfigure}[b]{0.3\linewidth}
		\includegraphics[width=\linewidth]{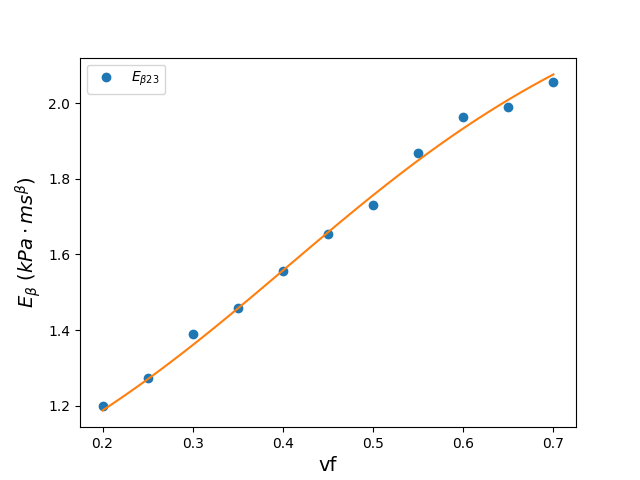}
		\caption{}
	\end{subfigure}
	
	\caption{Homogenized fractional viscoelastic spring-pot coefficients (generalized viscosity constant) as a function of volume fraction under six loading directions. Subfigures (a)–(f) correspond to loading directions 1-1, 2-2, 3-3, 1-2, 1-3, and 2-3, respectively.}
	\label{fig:E_beta}
\end{figure*}

\begin{figure*}[h!]
	\centering
	
	\begin{subfigure}[b]{0.3\linewidth}
		\includegraphics[width=\linewidth]{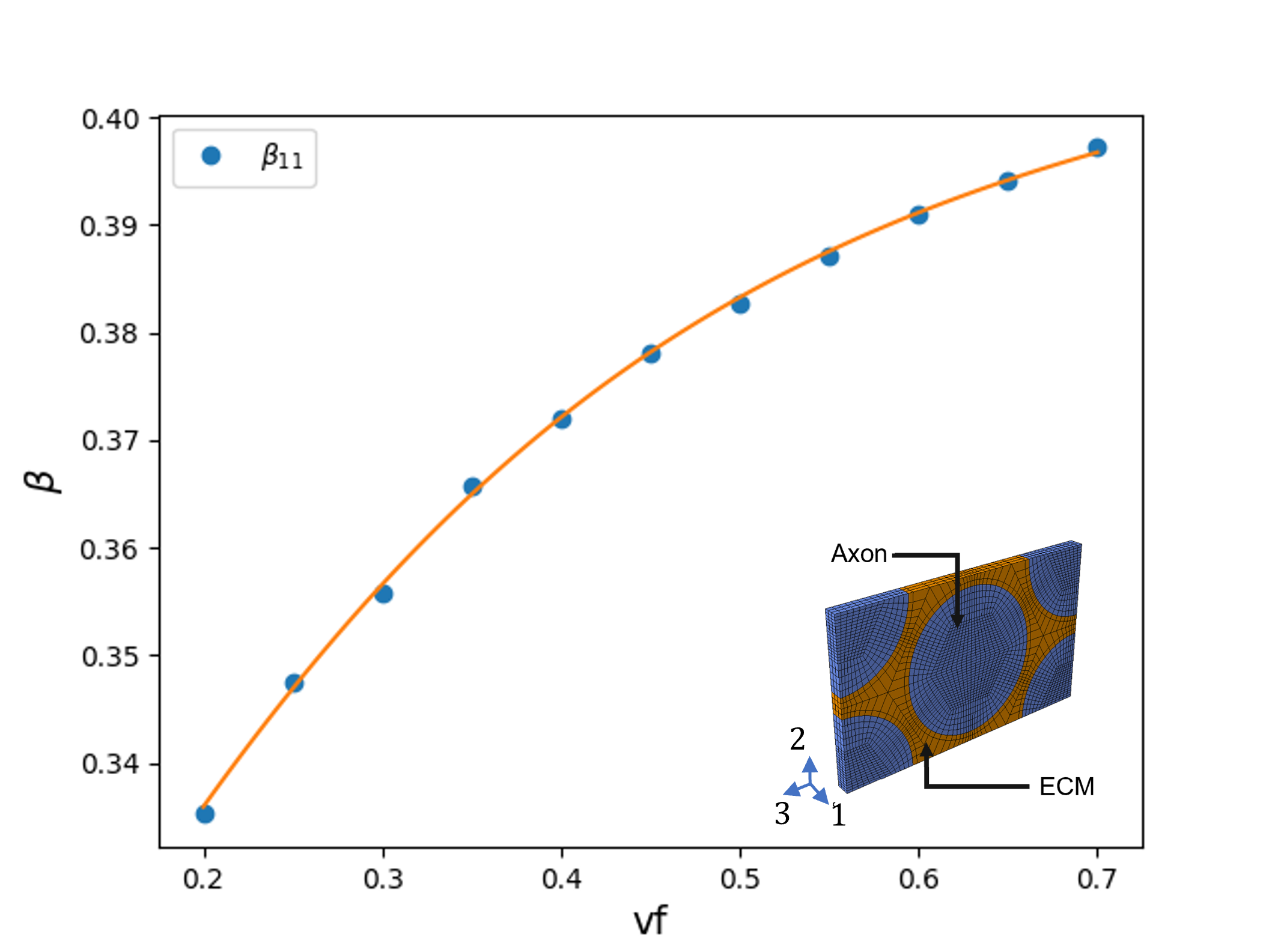}
		\caption{}
	\end{subfigure}
	\hfill
	\begin{subfigure}[b]{0.3\linewidth}
		\includegraphics[width=\linewidth]{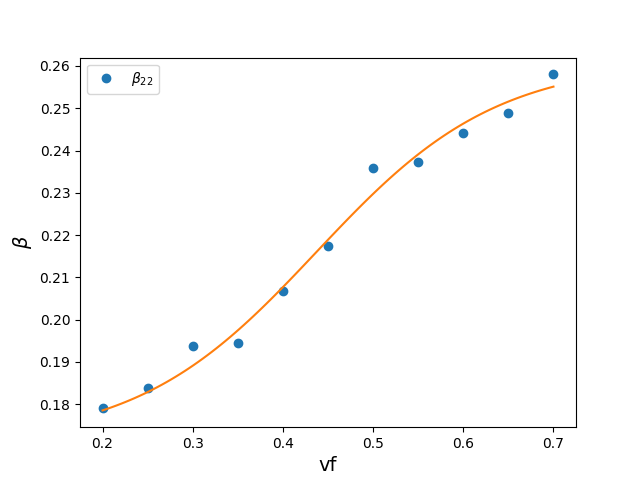}
		\caption{}
	\end{subfigure}
	\hfill
	\begin{subfigure}[b]{0.3\linewidth}
		\includegraphics[width=\linewidth]{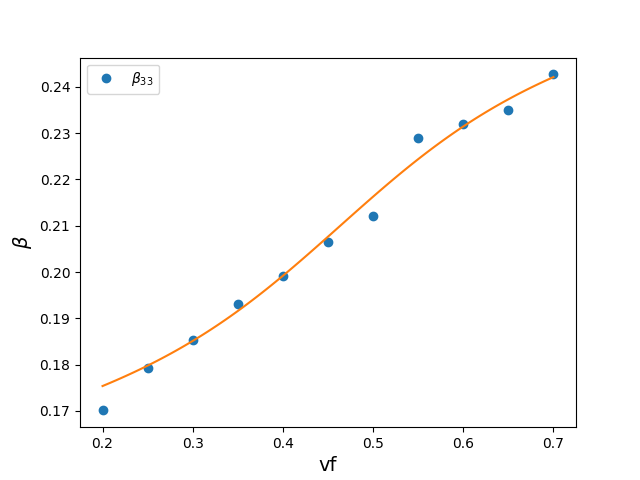}
		\caption{}
	\end{subfigure}
	
	\vspace{0.4cm}
	
	\begin{subfigure}[b]{0.3\linewidth}
		\includegraphics[width=\linewidth]{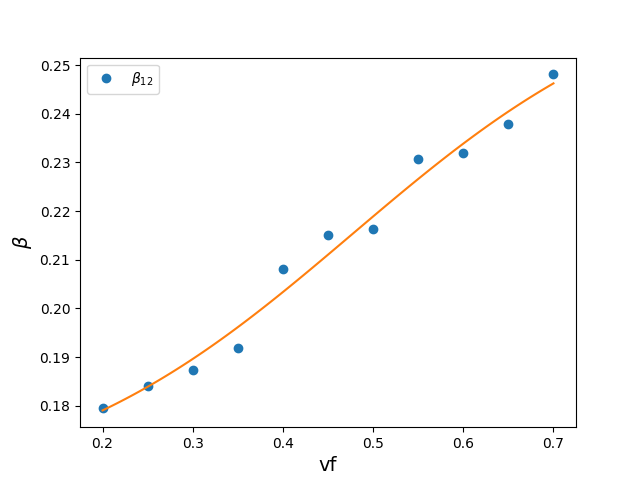}
		\caption{}
	\end{subfigure}
	\hfill
	\begin{subfigure}[b]{0.3\linewidth}
		\includegraphics[width=\linewidth]{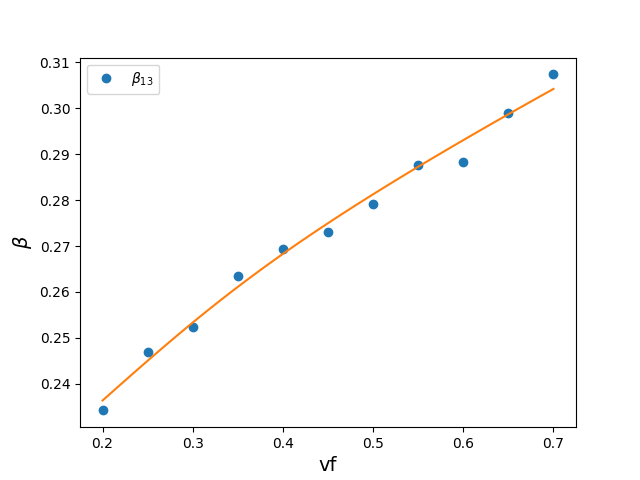}
		\caption{}
	\end{subfigure}
	\hfill
	\begin{subfigure}[b]{0.3\linewidth}
		\includegraphics[width=\linewidth]{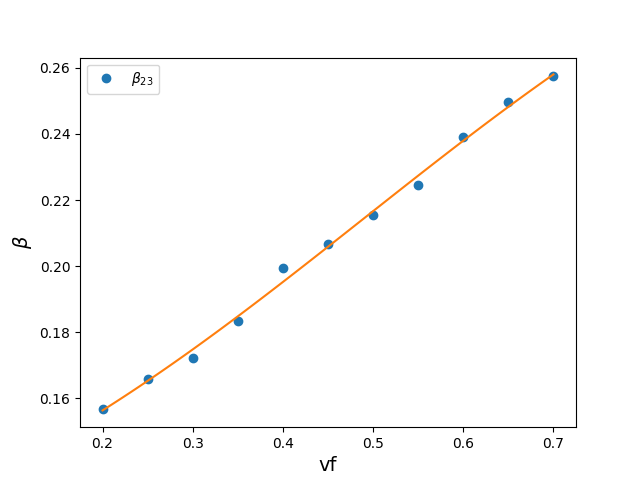}
		\caption{}
	\end{subfigure}
	
	\caption{Homogenized fractional viscoelastic spring-pot exponent as a function of volume fraction under six loading directions. Subfigures (a)–(f) correspond to loading directions 1-1, 2-2, 3-3, 1-2, 1-3, and 2-3, respectively.}
	\label{fig:beta}
\end{figure*}

\begin{figure*}[h]
	\centering
	
	\begin{subfigure}[b]{0.49\linewidth}
		\includegraphics[width=\linewidth]{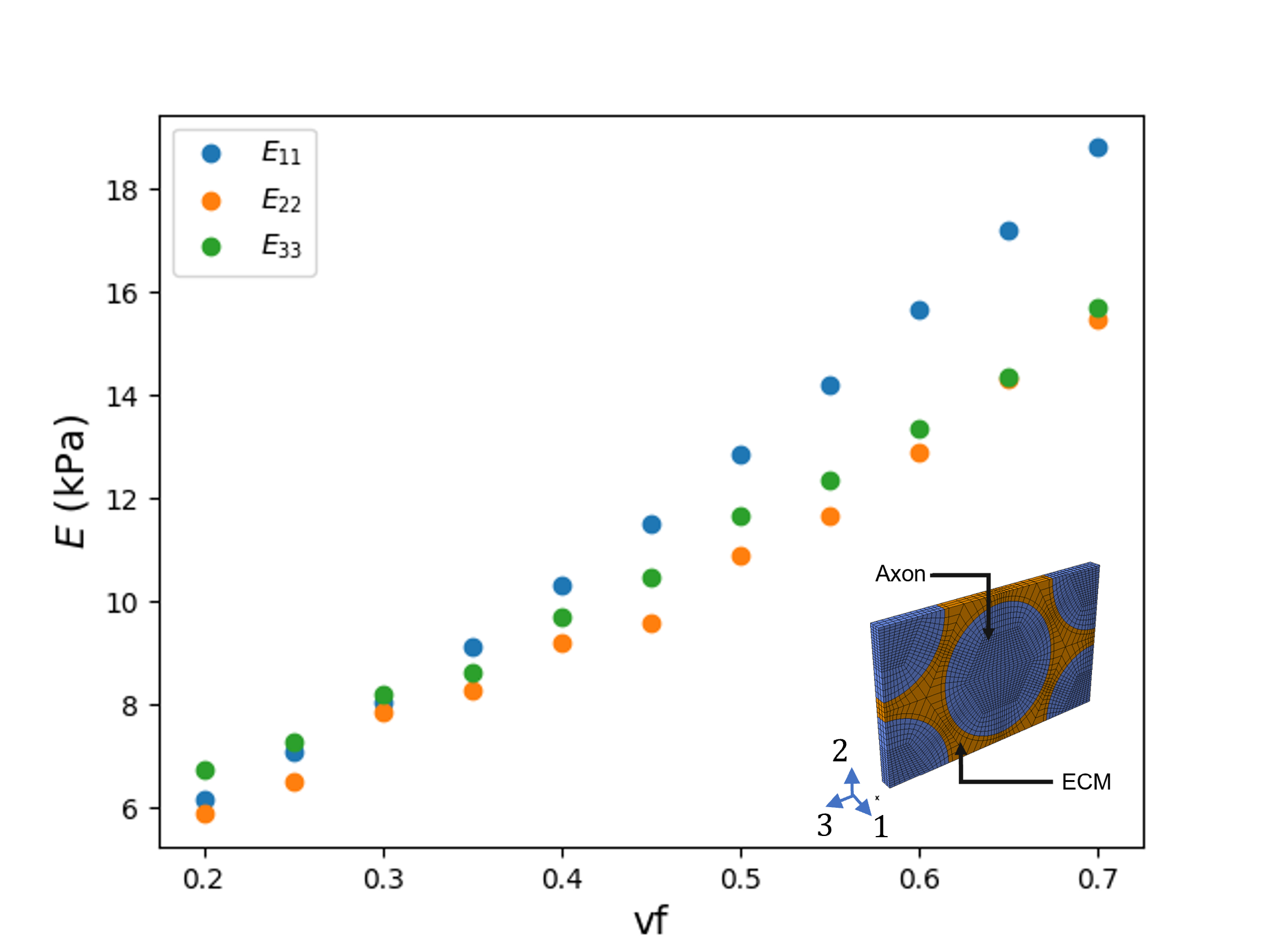}
		\caption{Elastic modulus along normal loading directions: 1-1, 2-2, and 3-3.}
		\label{fig:elastic_modulus_normal}
	\end{subfigure}
	\hfill
	\begin{subfigure}[b]{0.49\linewidth}
		\includegraphics[width=\linewidth]{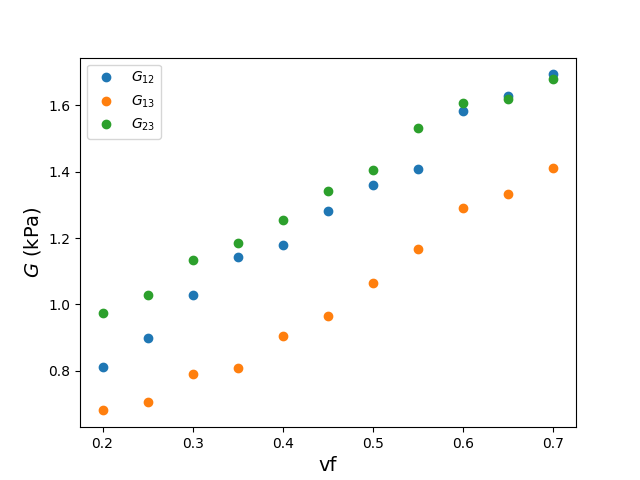}
		\caption{Elastic modulus along shear directions: 1-2, 1-3, and 2-3.}
		\label{fig:elastic_modulus_shear}
	\end{subfigure}
		\caption{Elastic moduli-like-parameters, with units $kPa$, under six loading directions obtained by combining the power-law parameters, $E_\beta$ and $\beta$.}
		\label{fig:elastic_modulus}
\end{figure*}

	The homogenized fractional viscoelastic parameters, $E_\beta$, $\beta$, of the RVE, as a function of the volume fraction, evaluated under six loading directions, are illustrated in \ref{fig:E_beta} and \ref{fig:beta}. $E_\beta$ (spring-pot coefficient) with units $(\frac{kPa}{ms^\beta})$, is known as the generalized viscosity constant, representing both, the material's stiffness as well as its viscous behavior. The power-law exponent $\beta$ serves as a bounding parameter in \cref{eq:eq_13} such that when $\beta = 0$, the model reduces to that of a spring with an elastic constant, whereas $\beta = 1$ produces the equation for a viscous dashpot. The two parameters taken together are used to compute the elastic modulus of the material. 
	
	The simulations indicate that the variation of the spring-pot coefficient, $E_\beta$, along the fiber direction is linear, whereas the variations along the other normal and shear directions are nonlinear. In the case of the fractional exponent $\beta$, its variation along the fiber direction follows a saturating exponential trend, approaching a plateau at high volume fractions. It can be observed that both $E_\beta$ and $\beta$ increase with higher axonal volume fractions. This trend aligns with existing data, which suggest that BWM glial cells are less viscous and significantly softer and more compliant than axons \cite{lu_viscoelastic_2006}. The linear variation of $E_{\beta_{11}}$ along the fiber direction is consistent with the rule-of-mixtures behavior observed in both strain rate–dependent \cite{pathan_effect_2017} and strain rate–independent \cite{banerjee_mechanical_2014, devireddy_effect_2014} fiber composite materials. The saturating exponential behavior of $\beta_{11}$ suggests that higher volume fractions lead to increased tissue viscosity.
	
	The homogenized viscoelastic parameters for the loading directions [2-2, 3-3, 1-2, 1-3, 2-3] were fit to various functional forms, with the best fit achieved using a converging bi-logistic function. The curve fit was obtained by solving a nonlinear least squares problem using the Trust Region Reflective algorithm implemented in SciPy's optimization library. This algorithm iteratively solves trust-region subproblems, where the objective function is locally approximated using the Gauss-Newton method. The size of the trust region is governed by both the proximity to the parameter bounds and the gradient direction. This method is particularly well-suited for bounded curve fitting problems. 
		
	\section{Discussion}
	\label{sec:Discussion}
	 
	Bi-logistic functions appear in the study of a wide range of biological systems \cite{meyer_bi-logistic_1994}. The bi-logistic fit reveals two distinct transitions (inflection points) in material properties as axon volume fraction changes. The rate of change of the homogenized properties shows two stiffening stages: a lower rate when the axon volume fraction is low, followed by a higher rate when larger axon volume fraction stiffens the RVE. Thus, the RVE transitions from a glial matrix-driven to an axon-driven mechanical response. The spring-pot coefficients $E_{\beta_{22}}$ and $E_{\beta_{33}}$ deviate from the rule of mixtures towards Halpin-Tsai models as expected in fiber reinforced composites. It is observed that RVE is slightly stiffer along the 3-3 ($E_{\beta_{33}}$) direction compared to the 2-2 direction ($E_{\beta_{22}}$). This can be attributed to the asymmetry of the RVE, which is more tightly packed vertically (2-2) than in the horizontal direction (3-3). This can also be observed under shearing with $E_{\beta_{12}}$ being greater than $E_{\beta_{13}}$.
		
	Given the units of $E_{\beta}$ $(\frac{kPa}{ms^\beta})$, the spring-pot parameters can be more  intuitively interpreted by combining them into a single shear-modulus-like parameter using the below equation \cite{sack_impact_2009}, 
	
	\begin{equation}
		E = (E_\beta * \eta^{-\beta})^{\frac{1}{1 - \beta}}, \hspace{3mm}
	\end{equation}

	\noindent where $E$ has the units of the shear modulus $(kPa)$ and $\eta = 3.7 \hspace{1mm}kPa \cdot ms$ is the viscosity parameter \cite{sack_impact_2009}. The elastic modulus along the three normal loading directions are shown in \Cref{fig:elastic_modulus_normal} and the modulus along the three shearing directions are shown in \Cref{fig:elastic_modulus_shear}. The results indicate that the variation of the elastic modulus with axon volume fraction is nonlinear in both the normal and shearing directions, and exhibits directional dependence---particularly in the CC, where axons are highly aligned and densely packed \cite{reiter_human_2025, hou_mechanical_2025}. This is particularly important in the context of neuropathological changes in different regions of the brain due to aging and disease. Multiple studies indicate that axon volume fraction does not remain constant with aging \cite{burzynska_correlates_2024, bouhrara_agerelated_2021}. It is therefore essential to quantify the effects of volume fraction on BWM stiffness, coupled with other factors such as myelin loss, axon degeneration, etc to be able to make robust neuropathological assessments. 
	
	The model predicts that the average elastic modulus along the fiber direction ($E_{11}$) for the CC is $12.9\ \text{kPa}$ which lies within the range of reported values for tensile modulus under quasi-static loading \cite{kang_systematic_2024, bernal_mechanical_2007, ouyang_contribution_2013}. The predicted value is also nearly twice the value reported by Kang et al. \cite{kang_viscoelastic_2024} in another study on porcine brain tissue under transverse compression loading. This effect is well documented in composite structures. Although fibers undergo mechanical stretch under both longitudinal tension and transverse compression, the mechanics differ. In the former, fibers carry load directly, leading to a stiffened response. In the latter, fibers are stretched indirectly through Poisson coupling, resulting in a matrix-dominated, weaker response. The model predictions are consistent with previous findings that brain tissue exhibits greater stiffness along the fiber direction than in the transverse direction \cite{atashgar_structure_2025}. This yields further support to the notion that axons are the primary contributors to the stiffness of the CC, despite the presence of other structural components such as blood vessels. While some studies report directional dependence of material properties in the CC \cite{reiter_human_2025, hou_mechanical_2025}, others have found no significant variation \cite{ kang_systematic_2024, budday_mechanical_2017}. The discrepancies among studies could stem from a range of additional factors, including differences in specimen type and size, temperature, experimental setup, and the nature of applied loading. 
	
	Along the shearing directions, the model predicts that the CC is stiffest under shear in the transverse direction ($G_{23}$), which is in agreement with existing experimental results \cite{arbogast1998material, reiter_human_2025, atashgar_structure_2025}. The average shear stress predicted by the model in CC is $1.11\ \text{kPa} - 1.5\ \text{kPa}$ which is in good agreement with a number of studies \cite{hou_mechanical_2025, bertalan_mechanical_2023, guo_towards_2013}. The model also predicts $G_{12} > G_{13}$. This again can be attributed to the asymmetry of hexagonal packing, with the fibers packed more closely vertically than horizontally. 
	
	The average power-law exponent under shear, predicted by the model is $0.2236 - 0.2531$. The values for $\beta$, reported for the whole brain varies from 0.265 \cite{sack_impact_2009, WUERFEL20102520}, to 0.339 \cite{kurt2019optimization}, and up to 0.91 \cite{morrison_atlas_2023}. For BWM, the reported power-law exponent values range from 0.337 \cite{kurt2019optimization} to  0.84 \cite{morrison_atlas_2023}. The atlas compiled by Morrison et al. \cite{morrison_atlas_2023} in their review article provides region-specific power-law values. For the CC, they report $\beta = 0.79$. This is considerably higher than $\beta$ reported in this porcine data-based study. While the above reported values were obtained from MRE studies of the human brain, the latter review reports higher values for both the whole brain and BWM. Along the axial direction, the predicted average value of $\beta$ is $0.38$ which is greater than the data from Kang et al ($0.216$) \cite{kang_viscoelastic_2024}, but both studies observed the highest $\beta$ along the fiber direction.
 	
	The significance of $\beta$ is that it encodes information about the underlying tissue architecture. In this study, this becomes evident through the evolution of $\beta$ as the volume fraction changes. This observation also offers a possible explanation for the variability of reported $\beta$ values across different studies i.e., the magnitude of $\beta$ varies based on the location and direction in which BWM is probed. A lower value of $\beta$ (closer to 0) represents a structure that is more solid whereas $\beta \approx 1$ is more representative of a fluid-like material dominated by viscous dissipation. From a rheological standpoint, the arrangement of springs and dashpots, representing distinct elastic and viscous effects can be tuned such that the overall response of the model reduces to the form in \cref{eq:eq_16}. Thus, the emergent parameter $\beta$ encapsulates the complex interplay between the different microstructural elements.

	\subsection{Efficient Computing}
	
	Given the computational expense of the GL operator, which requires storing and retrieving strain histories, we explore additional strategies to improve the computational efficiency of its implementation. For an explicit time integration scheme, in addition to the short-memory principle discussed in \Cref{subsec:mv}, increasing the time step size can improve the computing speed. However, the stable time increment depends on the dilatational wave speed of the material and the smallest element length. The above geometric and material constraints limit the size of the \enquote{stability limit}. Exceeding the stable time increment could result in numerical instability that may lead to unbounded solutions. With automatic solver-based time incrementation, even conservative estimates of the stable time increment can be increased as the simulation progresses and the element lengths change on account of the deformation. But simulations that require storing and retrieving field variables requires a fixed stable time increment which is computationally expensive. 
	
	Models with stored field variable histories implemented through user subroutines in Abaqus typically rely on \texttt{COMMON} block which define an area of memory accessible globally by all functions and subroutines in the code \cite{alotta2018finite, vestrum2015finite, liu2019dynamic, zheng2024application}. This, however, poses a number of challenges, especially during parallel execution. First, the definition of common blocks can vary between functions and subroutines which could lead to subtle errors with no obvious cause. Second, given the global scope of the common block, mutexes are essential to ensure thread safety during thread-based parallel execution. This, however, degrades performance. 
	
	A possible workaround to these challenges is to implement a Fortran 90–based \texttt{MODULE} for storing strain histories. Modern Fortran compilers including Intel and GNU support mixed F77-F90 codebases. Fortran modules enable explicit control over variable/array scope and promote modular code architecture. Furthermore, imports \texttt{(USE)} and exports \texttt{(PUBLIC)} are explicitly defined, which limits the scope of the strain history array to the module and the \texttt{VUMAT}, rather than being re-imported into every procedure or interface scope. Here we show that incorporating these methods result in thread-safe parallel code that yields significant time savings. 
	
\begin{figure}[h!]
		\centering
		\includegraphics[width=\columnwidth]{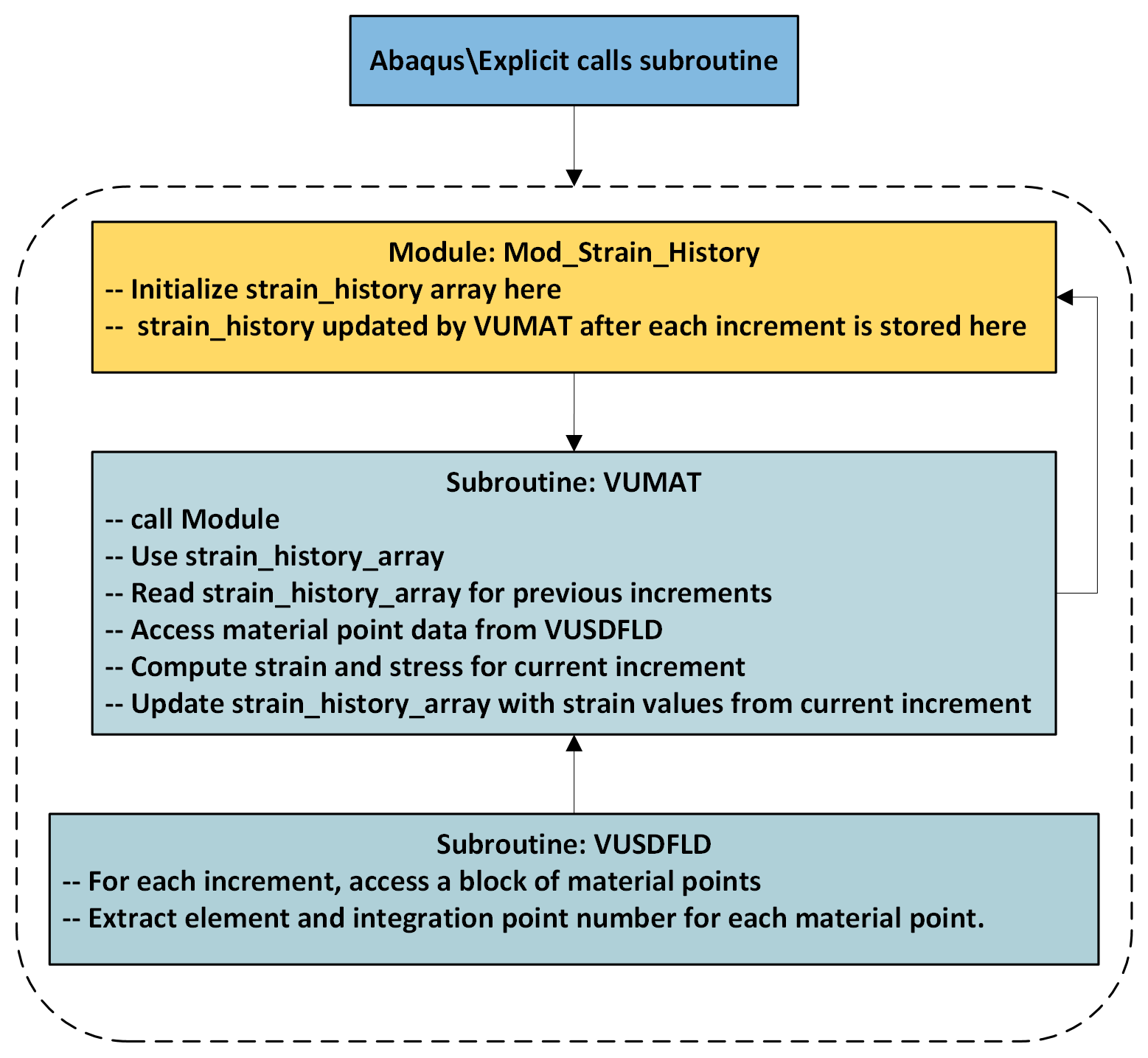}
		\caption{A representation of the algorithmic workflow for the \texttt{VUMAT}. The \texttt{VUSDFLD} subroutine tracks the material points at the start of each increment \cite{alotta2018finite}. The array that contains the strain history is defined and stored inside a module. The \texttt{VUMAT} subroutine calls \texttt{VUSDFLD}, the strain history array from the module, updates the strain history at the end of each increment for each element and integration point. The updated array is stored in the module.}
		\label{fig:Module_subroutine}
	\end{figure}

	In this study, we replace the common block methodology with a Fortran module that stores the strain histories (see \ref{fig:Module_subroutine}). The \texttt{VUMAT} subroutine calls another subroutine, \texttt{VUSDFLD},  which tracks the element number and integration point. The strain history array is then accessed by the \texttt{VUMAT} for retrieving the strains and computing the updated strain, which is stored back in the module. The module controls the visibility and access to the strain history array, thereby ensuring a more robust, thread-safe behavior during parallel execution. 
	
\begin{figure*}[htbp!]
		\centering
		\begin{subfigure}[b]{0.49\linewidth}
			\includegraphics[width=\linewidth]{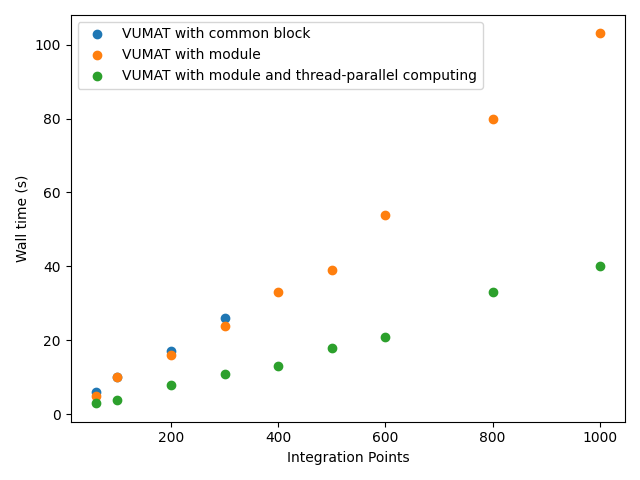}
			\caption{}
			\label{fig:Windows_runtime}
		\end{subfigure}
		\hfill
		\begin{subfigure}[b]{0.485\linewidth}
			\includegraphics[width=\linewidth]{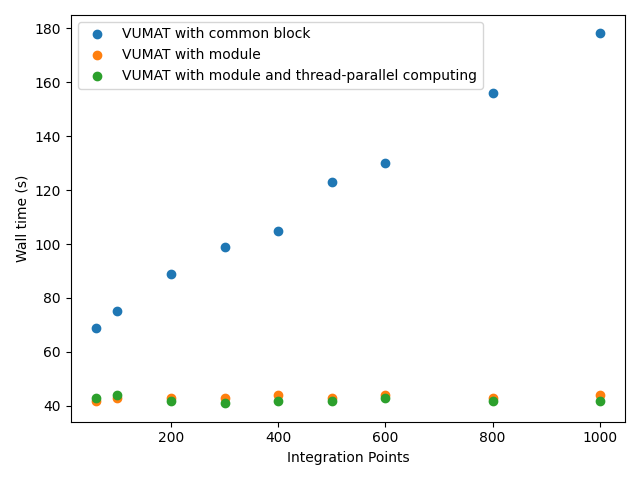}
			\caption{}
			\label{fig:Linux_runtime}
		\end
		{subfigure}
		\caption{A comparison of runtime performance between simulations using common-block, module and thread-parallel computing. (a) On a window machine with 16 GB RAM and 4 cores for parallel computing. (b) On a single node of a linux high performance cluster with 120 GB RAM and four cores for parallel computing. }
		\label{fig:runtimes}
	\end{figure*}

	The efficiency of a modular implementation of the \texttt{VUMAT} subroutine is demonstrated using a 1D beam example developed by Alota et al \cite{alotta2018finite}. The original example consists of a bar modeled with ten 1D linear truss elements, with its left end constrained and a $10N$ force applied at the right end, held constant for $t=5s$. We evaluate the performance of the subroutine by incrementally increasing the number of elements in the model and measuring the computation time. Since the strain history array must be allocated at the start of the subroutine, implementing it via common blocks results in large data structures being placed on the runtime stack, which can quickly exceed the stack size limit, causing execution failure. For the test case, with common blocks, on a windows machine with four cores and 16 GB RAM, the stack size limit was exceeded with 300 elements (see \Cref{fig:Windows_runtime}). On a linux high-performance cluster (HPC) that supports large data structures, using a module for the strain history in the \texttt{VUMAT} is significantly more efficient than using common blocks as the model size increases (see \Cref{fig:Linux_runtime}). 
	
	\begin{figure}[h]
	\centering
	\includegraphics[width=\columnwidth]{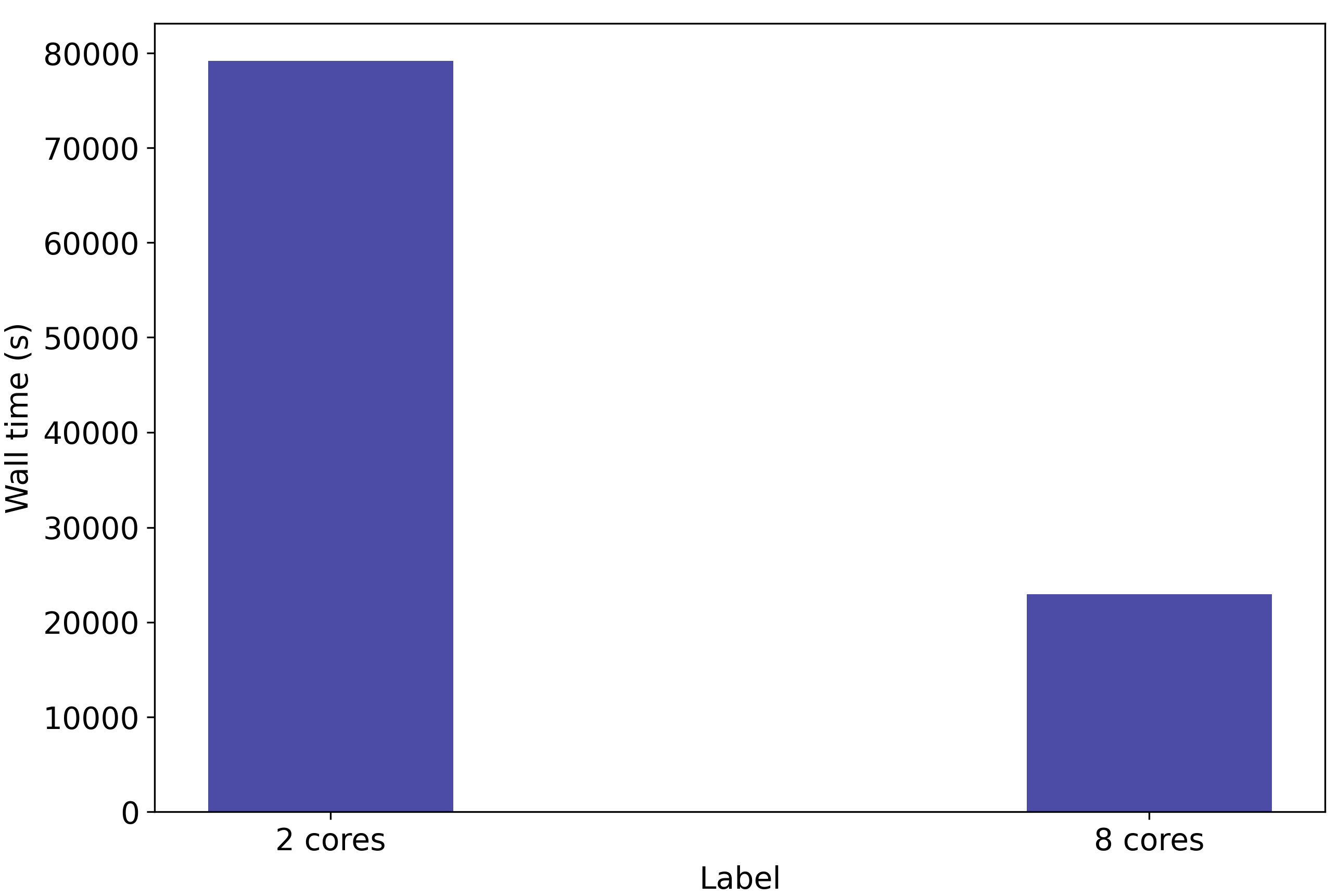}
	\caption{Runtime performance of the \texttt{VUMAT} subroutine evaluated using a plate-with-a-hole model consisting of 8,720 hexahedral elements with reduced integration and 50,000 explicit time increments. Simulations were performed on a single node with 120 GB RAM using 2 and 8 CPU cores. A 4× speedup is achieved, demonstrating good parallel scaling with core count.}
	\label{fig:plate_Parallel}
	\end{figure}

The performance of the two cases is compared with a thread-parallel execution of the \texttt{VUMAT} using the strain history module (see \ref{fig:runtimes}). On the windows machine with using 4 cores, the performance gain is visible even for small models such as the 1D bar example. On a Linux machine using 4 cores, the performance for the \texttt{VUMAT} with the module is similar between serial and parallel execution.  This is likely because, for a small model running on a large HPC, the serial code is already operating near its optimal capacity and therefore no real performance gain can be observed. Performance gains can be observed as the model size increases. For large 3D problems, computing performance can be further enhanced through shared-memory parallelism. The runtime performance for the plate-with-a-hole problem is evaluated on 2 and 8 CPU cores on a node with 120 GB RAM. The model achieves 4x speedup, demonstrating good scaling with core count. For the RVE with $vf=0.4$, the simulation time is reduced from 52 hours to 27 hours by using truncated memory and thread based parallelization on 16 cores and 80 GB RAM.  

	\subsection{Limitations}
	
		This study has a number of limitations. First, since BWM tissue exhibits dependence on both magnitude and rate of strain, the variation of homogenized properties with volume fraction obtained under small strain, quasi-static loading conditions may differ substantially under other loading profiles. Second, the model assumes a biphasic periodic RVE with uniformly distributed and perfectly aligned axons in an ECM. While some regions of BWM containing highly aligned, densely distributed axons can be approximated by such models, and such representations have been used in prior studies \cite{sullivan_sensitivity_2021, JAVID2014290, hasan_viscoelastic_2022}, they remain strictly idealized. In reality, even in regions such as the CC, axonal dispersion, waviness, and fiber crossings deviate markedly from this idealized representation. Furthermore, the impact of myelin which plays a vital role in regulating the mechanical response in BWM is not considered in this model. These factors highlight the need to move toward micromechanical models that account for architectural organization, i.e., clustering/random versus uniform distributions, presence of vascular structures, cellular tracts, etc \cite{reiter_human_2025}. 
		
		Third, the viscosity parameter used in this study is $\eta = 3.7\ \text{kPa} \cdot \text{ms}$, an average value defined for the whole white matter. To the best of our knowledge, consistent region-specific values for $\eta$ are not available. We acknowledge that accounting for volume-fraction-specific $\eta$ would alter the predicted elastic moduli. Also, since the extraction of homogenized properties is based on uniaxial tests derived from sparse material data, the resulting optimization problem is essentially unconstrained and multimodal. Consequently, multiple combinations of $E_\beta$ and $\beta$ can produce similar homogenized stress response. This issue is common in many calibration studies, where sparse experimental data and multimodal optimization landscapes make under/overfitting a challenge \cite{giudice_calibration_2021, babaei_machine_2022}. In order to force convergence to the global optimal set, one strategy is to set the starting parameter close to the global optimal set. However, this is only possible if the range of the admissible parameters is known and has a small size. An alternative approach is to increase the number of optimization iterations while constraining the allowed values of the variables. While this can improve the likelihood of finding the global minimum, it inevitably increases computational cost. This further underscores the need for larger data with multiple loading scenarios which would allow for physically motivated constraints in the optimization problem. 
	
		Finally, the fractional viscoelastic model provides a concise physical description of the observed power-law rheology of BWM compared to Prony series models. However, its implementation through the GL operator scheme is computationally expensive. Storing and retrieving field variable histories requires preallocating large amounts of memory, making the GL operator significantly more computationally expensive than Prony series models. This highlights the need to investigate the effects of further truncating the memory using circular binning for different loading regimes. Moving forward, investigations focusing on the implementation of faster algorithms, such as quadrature based solutions to Abel integral formulation for solving the RL fractional integral or sum of exponents approximations \cite{yuan_fast_2024} are essential.

	\section{Conclusions}
	
	To the best of our knowledge, this exploratory study is the first of its kind to propose and implement a 3D fractional viscoelastic finite element model of the corpus callosum of BWM in the time domain. Several soft tissues including BWM exhibit power-law responses under oscillatory shear tests such as MRE. In this study, drawing on the principle of scale invariance prominently expressed in the brain, we posit that the power-law behavior observed at the tissue level extends across length scales to the microstructural realm of the axons. This hypothesis has indeed formed the basis of multifrequency MRE \cite{sack2013structure}, but its application to cellular structures does not consider the  microstructural architecture and material anisotropy.

	It is often difficult to physically interpret the material response of such structures even in the linear limit of small deformations within the classical framework of springs and dashpots --- which give rise to Maxwell-Debye  exponential based Prony series. While the Prony series is readily implemented and widely available in commercial finite element software, it remains a strictly empirical representation that, at best, captures only a snapshot of the underlying dynamical processes. There are additional challenges; for example, even a seemingly simple two-parameter Prony series model requires the estimation of five material constants, and when incorporated into multiscale, multiphase models, the number of parameters can increase substantially. In this context, the language of fractional calculus, through the construct of a spring-pot provides a succinct ontological framework well-suited for modeling and  quantifying the power-law behavior observed in BWM. 
	
	The mechanical response of the homogenized tissue can be characterized by two parameters, namely a generalized viscosity constant, $E_\beta$, and a power-law exponent, $\beta$. $E_\beta$ represents the material's stiffness as well as its viscous behavior. $\beta$ serves as a bounding parameter which encodes information about the underlying tissue architecture. Using complex shear moduli data from oscillatory shear tests on porcine white matter tissue, a power-law model describing the material constants for the axons and ECM is developed. The fractional viscoelastic model is numerically implemented using the GL operator in a Fortran \texttt{VUMAT} subroutine and validated with benchmark examples. The biphasic finite element model of hexagonally packed axons in an ECM is developed in Abaqus FEA with periodic boundary conditions and a quasi-static displacement-controlled boundary condition. An explicit time integration scheme is used to compute the evolution of stress in the model along six loading directions for varying volume fractions of axons in an ECM. An optimization algorithm is then used to solve the inverse problem of extracting the homogenized properties for the biphasic model. 
	
	The model predicts that both $E_\beta$ and $\beta$ increase with higher axonal volume fractions indicating that glial cells are less viscous, significantly softer and more compliant than the axons. We also observe that the variation of $E_\beta$ with volume fraction, along the fiber direction is linear, whereas the variation along the transverse normal and shear directions are nonlinear. With $\beta$, the variation is nonlinear along all the loading directions with a saturating exponential behavior along the fiber direction. This indicates that $\beta$ increases rapidly at low volume fractions before saturating at high volume fractions with high viscosity. For off-fiber directions, the variation of both  $E_\beta$ and $\beta$ is best approximated  by a bi-logistic function with two stiffening stages: a lower, glial matrix-driven rate when the axon volume fraction is low, followed by a higher rate once larger axonal content reinforces the RVE. The magnitude of $E_\beta$ determines the solid-fluid behavior \cite{sack_impact_2009}.  The fractional viscoelastic model not only reinforces this interpretation but also sheds further light on the nature of its variation and directional dependence. By combining the fractional viscoelastic constants into a single shear-modulus-like parameter along each loading direction, we observe that the variation of elastic moduli with volume fraction along all loading directions is nonlinear. This nonlinearity is particularly important in the context of region-specific properties and neuropathological changes due to aging and disease. The model is well-suited for describing regions of BWM with aligned axonal fibers such as the CC. The predicted values of the elastic modulus both under tensile and shear loading are in good agreement with experimental observations.  
	
	In this study we also demonstrate an efficient, thread-safe implementation of the GL operator, which requires storing and retrieving of the strain history for computing the evolution of stress. The implementation leverages modern Fortran language based modules that enable explicit control over variable/array scope and promote modular code architecture. The performance gain in using a \texttt{Module}-based subroutine over traditional implementations with \texttt{COMMON} blocks is demonstrated through 1D and 3D examples on both, windows and linux computing platforms. We observe that the proposed methodology is significantly more efficient than existing approaches. Furthermore, by incorporating shared-memory parallelism, we achieve a 4x speedup in performance between 2 and 8 CPU cores. Overall, we demonstrate up to a $50 \%$ reduction in runtimes for the simulations of the BWM RVEs by incorporating modular array access, the short-memory principle and thread-parallel execution. 
	
	\section*{Ethics statement}
	The authors confirm that all procedures described in the paper were performed in compliance with relevant laws and institutional guidelines and have been approved by the appropriate institutional committee(s).
	
	\section*{Declaration of Competing Interest}
	The authors declare the following financial interests/personal relationships which may be considered as potential competing interests:
	
	Assimina A Pelegri reports financial support was provided by the National Science Foundation. John G Georgiadis reports financial support from the  R.A. Pritzker endowed chair. Parameshwaran Pasupathy has no known competing financial interests or personal relationships that could have appeared to influence the work reported in this paper.
	
	\section*{Acknowledgments}
	The authors gratefully acknowledge the support provided by NSF via CMMI-1436743, CMMI-1762774 grants and by the R.A. Pritzker endowed chair. The authors also gratefully acknowledge Baoxiang Shan, PhD., for insightful discussions on the development of Fortran subroutines. In addition, the authors also thank Stephen S. Recchia, Ph.D., and Adrian Blot for their valuable insights on debugging and improving code performance. Finally, the authors thank Nader Fateh of ESTECO North America for generously granting a research modeFRONTIER license. 
	\bibliographystyle{elsarticle-num}
	\bibliography{fractional_model_ref}

\begin{thebibliography}{10}
\expandafter\ifx\csname url\endcsname\relax
  \def\url#1{\texttt{#1}}\fi
\expandafter\ifx\csname urlprefix\endcsname\relax\def\urlprefix{URL }\fi
\expandafter\ifx\csname href\endcsname\relax
  \def\href#1#2{#2} \def\path#1{#1}\fi

\bibitem{bullock_taxonomy_2022}
D.~N. Bullock, E.~A. Hayday, M.~D. Grier, W.~Tang, F.~Pestilli, S.~R.
  Heilbronner,
  \href{https://academic.oup.com/cercor/article/32/20/4524/6528900}{A taxonomy
  of the brain’s white matter: twenty-one major tracts for the 21st century},
  Cerebral Cortex 32~(20) (2022) 4524--4548.
\newblock \href {https://doi.org/10.1093/cercor/bhab500}
  {\path{doi:10.1093/cercor/bhab500}}.

\bibitem{chatelin_fifty_2010}
S.~Chatelin, A.~Constantinesco, R.~Willinger,
  \href{https://www.medra.org/servlet/aliasResolver?alias=iospress&doi=10.3233/BIR-2010-0576}{Fifty
  years of brain tissue mechanical testing: {From} in vitro to in vivo
  investigations}, Biorheology 47~(5-6) (2010) 255--276.
\newblock \href {https://doi.org/10.3233/BIR-2010-0576}
  {\path{doi:10.3233/BIR-2010-0576}}.

\bibitem{brainsci14040401}
Y.-J. Jeon, S.-E. Park, H.-M. Baek,
  \href{https://www.mdpi.com/2076-3425/14/4/401}{Predicting brain age and
  gender from brain volume data using variational quantum circuits}, Brain
  Sciences 14~(4) (2024).
\newblock \href {https://doi.org/10.3390/brainsci14040401}
  {\path{doi:10.3390/brainsci14040401}}.

\bibitem{sack_impact_2009}
I.~Sack, B.~Beierbach, J.~Wuerfel, D.~Klatt, U.~Hamhaber, S.~Papazoglou,
  P.~Martus, J.~Braun, The impact of aging and gender on brain viscoelasticity,
  Neuroimage 46~(3) (2009) 652--657, publisher: Elsevier.

\bibitem{burzynska_correlates_2024}
A.~Z. Burzynska, C.~Anderson, D.~B. Arciniegas, V.~Calhoun, I.-Y. Choi,
  A.~Mendez~Colmenares, A.~F. Kramer, K.~Li, J.~Lee, P.~Lee, M.~L. Thomas,
  \href{https://linkinghub.elsevier.com/retrieve/pii/S2666245024000047}{Correlates
  of axonal content in healthy adult span: {Age}, sex, myelin, and metabolic
  health}, Cerebral Circulation - Cognition and Behavior 6 (2024) 100203.
\newblock \href {https://doi.org/10.1016/j.cccb.2024.100203}
  {\path{doi:10.1016/j.cccb.2024.100203}}.

\bibitem{arbogast1998material}
K.~B. Arbogast, S.~S. Margulies, Material characterization of the brainstem
  from oscillatory shear tests, Journal of biomechanics 31~(9) (1998) 801--807.

\bibitem{SAIKI2009549}
R.~L. Saiki,
  \href{https://www.sciencedirect.com/science/article/pii/S0899588509000392}{Current
  and evolving management of traumatic brain injury}, Critical Care Nursing
  Clinics of North America 21~(4) (2009) 549--559, neuroscience.
\newblock \href {https://doi.org/https://doi.org/10.1016/j.ccell.2009.07.009}
  {\path{doi:https://doi.org/10.1016/j.ccell.2009.07.009}}.

\bibitem{suzuki_estimation_2016}
Y.~Suzuki, M.~Hori, K.~Kamiya, I.~Fukunaga, S.~Aoki, M.~Van~Cauteren,
  \href{https://www.jstage.jst.go.jp/article/mrms/15/1/15_2014-0141/_article}{Estimation
  of the {Mean} {Axon} {Diameter} and {Intra}-axonal {Space} {Volume}
  {Fraction} of the {Human} {Corpus} {Callosum}: {Diffusion} q-space {Imaging}
  with {Low} q-values}, MRMS 15~(1) (2016) 83--93.
\newblock \href {https://doi.org/10.2463/mrms.2014-0141}
  {\path{doi:10.2463/mrms.2014-0141}}.

\bibitem{fan_age-related_2019}
Q.~Fan, Q.~Tian, N.~A. Ohringer, A.~Nummenmaa, T.~Witzel, S.~M. Tobyne, E.~C.
  Klawiter, C.~Mekkaoui, B.~R. Rosen, L.~L. Wald, D.~H. Salat, S.~Y. Huang,
  \href{https://linkinghub.elsevier.com/retrieve/pii/S1053811919301302}{Age-related
  alterations in axonal microstructure in the corpus callosum measured by
  high-gradient diffusion {MRI}}, NeuroImage 191 (2019) 325--336.
\newblock \href {https://doi.org/10.1016/j.neuroimage.2019.02.036}
  {\path{doi:10.1016/j.neuroimage.2019.02.036}}.

\bibitem{eilbes2025neuroanatomical}
M.~Eilbes, Neuroanatomical planes in human brains,
  \url{https://app.biorender.com/biorender-templates/details/t-6528951793007873fb7fa81a-neuroanatomical-planes-in-human-brains},
  accessed: 2025-08-08 (2025).

\bibitem{edwards_microstructural_2024}
T.~J. Edwards, R.~J. Dean, G.~A. Robinson, J.~Knight, S.~A. Mandelstam, L.~J.
  Richards,
  \href{https://linkinghub.elsevier.com/retrieve/pii/S2213158224001335}{Microstructural
  analysis does not support altered interhemispheric wiring of the human
  anterior commissure in corpus callosum dysgenesis}, NeuroImage: Clinical 44
  (2024) 103692.
\newblock \href {https://doi.org/10.1016/j.nicl.2024.103692}
  {\path{doi:10.1016/j.nicl.2024.103692}}.

\bibitem{huang_high-gradient_2020}
S.~Y. Huang, Q.~Tian, Q.~Fan, T.~Witzel, B.~Wichtmann, J.~A. McNab,
  J.~Daniel~Bireley, N.~Machado, E.~C. Klawiter, C.~Mekkaoui, L.~L. Wald,
  A.~Nummenmaa,
  \href{http://link.springer.com/10.1007/s00429-019-01961-2}{High-gradient
  diffusion {MRI} reveals distinct estimates of axon diameter index within
  different white matter tracts in the in vivo human brain}, Brain Struct Funct
  225~(4) (2020) 1277--1291.
\newblock \href {https://doi.org/10.1007/s00429-019-01961-2}
  {\path{doi:10.1007/s00429-019-01961-2}}.

\bibitem{johnson_local_2013}
C.~L. Johnson, M.~D. McGarry, A.~A. Gharibans, J.~B. Weaver, K.~D. Paulsen,
  H.~Wang, W.~C. Olivero, B.~P. Sutton, J.~G. Georgiadis,
  \href{https://linkinghub.elsevier.com/retrieve/pii/S1053811913004485}{Local
  mechanical properties of white matter structures in the human brain},
  NeuroImage 79 (2013) 145--152.
\newblock \href {https://doi.org/10.1016/j.neuroimage.2013.04.089}
  {\path{doi:10.1016/j.neuroimage.2013.04.089}}.

\bibitem{WUERFEL20102520}
J.~Wuerfel, F.~Paul, B.~Beierbach, U.~Hamhaber, D.~Klatt, S.~Papazoglou,
  F.~Zipp, P.~Martus, J.~Braun, I.~Sack,
  \href{https://www.sciencedirect.com/science/article/pii/S1053811909006338}{Mr-elastography
  reveals degradation of tissue integrity in multiple sclerosis}, NeuroImage
  49~(3) (2010) 2520--2525.
\newblock \href
  {https://doi.org/https://doi.org/10.1016/j.neuroimage.2009.06.018}
  {\path{doi:https://doi.org/10.1016/j.neuroimage.2009.06.018}}.

\bibitem{kiss_low-frequency_2024}
C.~Kiss, S.~Wurth, B.~Heschl, M.~Khalil, T.~Gattringer, C.~Enzinger, S.~Ropele,
  \href{https://linkinghub.elsevier.com/retrieve/pii/S2213158224000457}{Low-frequency
  {MR} elastography reveals altered deep gray matter viscoelasticity in
  multiple sclerosis}, NeuroImage: Clinical 42 (2024) 103606.
\newblock \href {https://doi.org/10.1016/j.nicl.2024.103606}
  {\path{doi:10.1016/j.nicl.2024.103606}}.

\bibitem{sullivan_sensitivity_2021}
D.~J. Sullivan, X.~Wu, N.~R. Gallo, N.~M. Naughton, J.~G. Georgiadis, A.~A.
  Pelegri, Sensitivity analysis of effective transverse shear viscoelastic and
  diffusional properties of myelinated white matter, Physics in Medicine and
  Biology 66 (2021) 3.

\bibitem{montanino2018}
A.~Montanino, S.~Kleiven, Utilizing a structural mechanics approach to assess
  the primary effects of injury loads onto the axon and its components,
  Frontiers in neurology (2018) 643.

\bibitem{JAVID2014290}
S.~Javid, A.~Rezaei, G.~Karami,
  \href{https://www.sciencedirect.com/science/article/pii/S1751616113003986}{A
  micromechanical procedure for viscoelastic characterization of the axons and
  ecm of the brainstem}, Journal of the Mechanical Behavior of Biomedical
  Materials 30 (2014) 290--299.
\newblock \href {https://doi.org/https://doi.org/10.1016/j.jmbbm.2013.11.010}
  {\path{doi:https://doi.org/10.1016/j.jmbbm.2013.11.010}}.

\bibitem{reiter_modeling_2023}
N.~Reiter, A.~Schäfer, S.~Auer, F.~Paulsen, S.~Budday,
  \href{https://onlinelibrary.wiley.com/doi/10.1002/pamm.202300234}{Modeling
  the finite viscoelasticity of human brain tissue based on microstructural
  information}, Proc Appl Math and Mech 23~(4) (2023) e202300234.
\newblock \href {https://doi.org/10.1002/pamm.202300234}
  {\path{doi:10.1002/pamm.202300234}}.

\bibitem{li_investigation_2021}
W.~Li, D.~E.~T. Shepherd, D.~M. Espino,
  \href{https://link.springer.com/10.1007/s10439-021-02866-0}{Investigation of
  the {Compressive} {Viscoelastic} {Properties} of {Brain} {Tissue} {Under}
  {Time} and {Frequency} {Dependent} {Loading} {Conditions}}, Ann Biomed Eng
  49~(12) (2021) 3737--3747.
\newblock \href {https://doi.org/10.1007/s10439-021-02866-0}
  {\path{doi:10.1007/s10439-021-02866-0}}.

\bibitem{lai_machine_2020}
C.~Lai, Y.~Chen, T.~Wang, J.~Liu, Q.~Wang, Y.~Du, Y.~Feng,
  \href{https://link.springer.com/10.1007/s11517-020-02262-1}{A machine
  learning approach for magnetic resonance image–based mouse brain modeling
  and fast computation in controlled cortical impact}, Med Biol Eng Comput
  58~(11) (2020) 2835--2844.
\newblock \href {https://doi.org/10.1007/s11517-020-02262-1}
  {\path{doi:10.1007/s11517-020-02262-1}}.

\bibitem{wu_application_2024}
X.~Wu, P.~Pasupathy, A.~A. Pelegri,
  \href{https://linkinghub.elsevier.com/retrieve/pii/S0169260724003742}{On the
  application of hybrid deep {3D} convolutional neural network algorithms for
  predicting the micromechanics of brain white matter}, Computer Methods and
  Programs in Biomedicine 256 (2024) 108381.
\newblock \href {https://doi.org/10.1016/j.cmpb.2024.108381}
  {\path{doi:10.1016/j.cmpb.2024.108381}}.

\bibitem{sack2013structure}
I.~Sack, K.~J{\"o}hrens, J.~W{\"u}rfel, J.~Braun, Structure-sensitive
  elastography: on the viscoelastic powerlaw behavior of in vivo human tissue
  in health and disease, Soft matter 9~(24) (2013) 5672--5680.

\bibitem{fehlner_higherresolution_2016}
A.~Fehlner, J.~R. Behrens, K.~Streitberger, S.~Papazoglou, J.~Braun,
  J.~Bellmann‐Strobl, K.~Ruprecht, F.~Paul, J.~Würfel, I.~Sack,
  \href{https://onlinelibrary.wiley.com/doi/10.1002/jmri.25129}{Higher‐resolution
  {MR} elastography reveals early mechanical signatures of neuroinflammation in
  patients with clinically isolated syndrome}, Magnetic Resonance Imaging
  44~(1) (2016) 51--58.
\newblock \href {https://doi.org/10.1002/jmri.25129}
  {\path{doi:10.1002/jmri.25129}}.

\bibitem{kurt2019optimization}
M.~Kurt, L.~Wu, K.~Laksari, E.~Ozkaya, Z.~M. Suar, H.~Lv, K.~Epperson,
  K.~Epperson, A.~M. Sawyer, D.~Camarillo, et~al., Optimization of a
  multifrequency magnetic resonance elastography protocol for the human brain,
  Journal of Neuroimaging 29~(4) (2019) 440--446.

\bibitem{nicolas2018biomechanical}
E.~Nicolas, S.~Calle, S.~Nicolle, D.~Mitton, J.-P. Remenieras, Biomechanical
  characterization of ex vivo human brain using ultrasound shear wave
  spectroscopy, Ultrasonics 84 (2018) 119--125.

\bibitem{kang_viscoelastic_2024}
W.~Kang, L.~Wang, Y.~Fan,
  \href{https://linkinghub.elsevier.com/retrieve/pii/S0021929023004591}{Viscoelastic
  response of gray matter and white matter brain tissues under creep and
  relaxation}, Journal of Biomechanics 162 (2024) 111888.
\newblock \href {https://doi.org/10.1016/j.jbiomech.2023.111888}
  {\path{doi:10.1016/j.jbiomech.2023.111888}}.

\bibitem{mishra_rheological_2024}
A.~Mishra, R.~O. Cleveland,
  \href{https://www.frontiersin.org/articles/10.3389/fbioe.2024.1386955/full}{Rheological
  properties of porcine organs: measurements and fractional viscoelastic
  model}, Front. Bioeng. Biotechnol. 12 (2024) 1386955.
\newblock \href {https://doi.org/10.3389/fbioe.2024.1386955}
  {\path{doi:10.3389/fbioe.2024.1386955}}.

\bibitem{grosu_fractal_2023}
G.~F. Grosu, A.~V. Hopp, V.~V. Moca, H.~Bârzan, A.~Ciuparu, M.~Ercsey-Ravasz,
  M.~Winkel, H.~Linde, R.~C. Mureșan,
  \href{https://academic.oup.com/cercor/article/33/8/4574/6713293}{The fractal
  brain: scale-invariance in structure and dynamics}, Cerebral Cortex 33~(8)
  (2023) 4574--4605.
\newblock \href {https://doi.org/10.1093/cercor/bhac363}
  {\path{doi:10.1093/cercor/bhac363}}.

\bibitem{pasupathy_fractional_2023}
P.~Pasupathy, J.~G. Georgiadis, A.~A. Pelegri,
  \href{http://aip.scitation.org/doi/abs/10.1063/5.0163389}{A fractional
  viscoelastic model of the axon in brain white matter}, Belgrade, Serbia,
  2023, p. 070002.
\newblock \href {https://doi.org/10.1063/5.0163389}
  {\path{doi:10.1063/5.0163389}}.

\bibitem{arbogast_fiber-reinforced_1999}
K.~B. Arbogast, S.~S. Margulies,
  \href{https://linkinghub.elsevier.com/retrieve/pii/S0021929099000421}{A
  fiber-reinforced composite model of the viscoelastic behavior of the
  brainstem in shear}, Journal of Biomechanics 32~(8) (1999) 865--870,
  publisher: Elsevier BV.
\newblock \href {https://doi.org/10.1016/s0021-9290(99)00042-1}
  {\path{doi:10.1016/s0021-9290(99)00042-1}}.

\bibitem{WebPlotDigitizer}
A.~Rohatgi, \href{https://automeris.io}{Webplotdigitizer}.

\bibitem{caffyn_significance_1945}
J.~E. Caffyn, G.~W.~S. Blair,
  \href{https://www.nature.com/articles/155171c0}{Significance of {Power}-{Law}
  {Relations} in {Rheology}}, Nature 155~(3928) (1945) 171--172.
\newblock \href {https://doi.org/10.1038/155171c0}
  {\path{doi:10.1038/155171c0}}.

\bibitem{scott_blair_role_1947}
G.~Scott~Blair,
  \href{https://linkinghub.elsevier.com/retrieve/pii/009585224790007X}{The role
  of psychophysics in rheology}, Journal of Colloid Science 2~(1) (1947)
  21--32.
\newblock \href {https://doi.org/10.1016/0095-8522(47)90007-X}
  {\path{doi:10.1016/0095-8522(47)90007-X}}.

\bibitem{noauthor_survey_1999}
\href{https://linkinghub.elsevier.com/retrieve/pii/S0076539299800290}{Survey of
  {Applications} of the {Fractional} {Calculus}}, in: Mathematics in {Science}
  and {Engineering}, Vol. 198, Elsevier, 1999, pp. 261--307.
\newblock \href {https://doi.org/10.1016/S0076-5392(99)80029-0}
  {\path{doi:10.1016/S0076-5392(99)80029-0}}.

\bibitem{fedorchenko_introduction_2008}
A.~I. Fedorchenko,
  \href{http://rgdoi.net/10.13140/2.1.1890.6887}{{INTRODUCTION} {TO}
  {FRACTIONAL} {CALCULUS}} (2008).
\newblock \href {https://doi.org/10.13140/2.1.1890.6887}
  {\path{doi:10.13140/2.1.1890.6887}}.

\bibitem{bagley_theoretical_1983}
R.~L. Bagley, P.~J. Torvik,
  \href{https://pubs.aip.org/sor/jor/article/27/3/201-210/234686}{A
  {Theoretical} {Basis} for the {Application} of {Fractional} {Calculus} to
  {Viscoelasticity}}, Journal of Rheology 27~(3) (1983) 201--210.
\newblock \href {https://doi.org/10.1122/1.549724}
  {\path{doi:10.1122/1.549724}}.

\bibitem{bonfanti_fractional_2020}
A.~Bonfanti, J.~L. Kaplan, G.~Charras, A.~Kabla,
  \href{https://xlink.rsc.org/?DOI=D0SM00354A}{Fractional viscoelastic models
  for power-law materials}, Soft Matter 16~(26) (2020) 6002--6020.
\newblock \href {https://doi.org/10.1039/D0SM00354A}
  {\path{doi:10.1039/D0SM00354A}}.

\bibitem{alotta2017behavior}
G.~Alotta, O.~Barrera, A.~C. Cocks, M.~D. Paola, On the behavior of a
  three-dimensional fractional viscoelastic constitutive model, Meccanica
  52~(9) (2017) 2127--2142.

\bibitem{199941}
\href{https://www.sciencedirect.com/science/article/pii/S0076539299800216}{Chapter
  2 - fractional derivatives and integrals}, in: I.~Podlubny (Ed.), Fractional
  Differential Equations, Vol. 198 of Mathematics in Science and Engineering,
  Elsevier, 1999, pp. 41--119.
\newblock \href {https://doi.org/https://doi.org/10.1016/S0076-5392(99)80021-6}
  {\path{doi:https://doi.org/10.1016/S0076-5392(99)80021-6}}.

\bibitem{schmidt_no_2002}
A.~Schmidt, L.~Gaul,
  \href{http://link.springer.com/10.1023/A:1016552503411}{Finite element
  formulation of viscoelastic constitutive equations using fractional time
  derivatives}, Nonlinear Dynamics 29~(1/4) (2002) 37--55.
\newblock \href {https://doi.org/10.1023/A:1016552503411}
  {\path{doi:10.1023/A:1016552503411}}.

\bibitem{alotta2018finite}
G.~Alotta, O.~Barrera, A.~Cocks, M.~Di~Paola, The finite element implementation
  of 3d fractional viscoelastic constitutive models, Finite Elements in
  Analysis and Design 146 (2018) 28--41.

\bibitem{1999199}
\href{https://www.sciencedirect.com/science/article/pii/S0076539299800265}{Chapter
  7 - numerical evaluation of fractional derivatives}, in: I.~Podlubny (Ed.),
  Fractional Differential Equations, Vol. 198 of Mathematics in Science and
  Engineering, Elsevier, 1999, pp. 199--221.
\newblock \href {https://doi.org/https://doi.org/10.1016/S0076-5392(99)80026-5}
  {\path{doi:https://doi.org/10.1016/S0076-5392(99)80026-5}}.

\bibitem{tian2019periodic}
W.~Tian, L.~Qi, X.~Chao, J.~Liang, M.~Fu, Periodic boundary condition and its
  numerical implementation algorithm for the evaluation of effective mechanical
  properties of the composites with complicated micro-structures, Composites
  Part B: Engineering 162 (2019) 1--10.

\bibitem{sadaba_special-purpose_2019}
S.~Sádaba, M.~Herráez, F.~Naya, C.~González, J.~Llorca, C.~Lopes,
  \href{https://linkinghub.elsevier.com/retrieve/pii/S0263822318328186}{Special-purpose
  elements to impose {Periodic} {Boundary} {Conditions} for multiscale
  computational homogenization of composite materials with the explicit
  {Finite} {Element} {Method}}, Composite Structures 208 (2019) 434--441.
\newblock \href {https://doi.org/10.1016/j.compstruct.2018.10.037}
  {\path{doi:10.1016/j.compstruct.2018.10.037}}.

\bibitem{abq}
M.~Smith, ABAQUS/Standard User's Manual, Version 2020, Dassault Syst{\`e}mes
  Simulia Corp, United States, 2020.

\bibitem{omairey2019development}
S.~L. Omairey, P.~D. Dunning, S.~Sriramula, Development of an abaqus plugin
  tool for periodic rve homogenisation, Engineering with Computers 35~(2)
  (2019) 567--577.

\bibitem{hesammokri2019implementation}
P.~Hesammokri, Implementation of fractional order viscoelastic models to finite
  element method, Master's thesis, Middle East Technical University (2019).

\bibitem{andersson2012derivative}
C.~Andersson, S.~Gedda, J.~{\AA}kesson, S.~Diehl, Derivative-free parameter
  optimization of functional mock-up units, in: 9th International Modelica
  Conference, Modelica Association, 2012.

\bibitem{mf}
ESTECO, modefrontier - multi objective optimization design environment,
  \url{www.esteco.com} (2022).

\bibitem{lu_viscoelastic_2006}
Y.-B. Lu, K.~Franze, G.~Seifert, C.~Steinhäuser, F.~Kirchhoff, H.~Wolburg,
  J.~Guck, P.~Janmey, E.-Q. Wei, J.~Käs, A.~Reichenbach,
  \href{https://pnas.org/doi/full/10.1073/pnas.0606150103}{Viscoelastic
  properties of individual glial cells and neurons in the {CNS}}, Proc. Natl.
  Acad. Sci. U.S.A. 103~(47) (2006) 17759--17764.
\newblock \href {https://doi.org/10.1073/pnas.0606150103}
  {\path{doi:10.1073/pnas.0606150103}}.

\bibitem{pathan_effect_2017}
M.~Pathan, V.~Tagarielli, S.~Patsias,
  \href{https://linkinghub.elsevier.com/retrieve/pii/S0263822316318281}{Effect
  of fibre shape and interphase on the anisotropic viscoelastic response of
  fibre composites}, Composite Structures 162 (2017) 156--163.
\newblock \href {https://doi.org/10.1016/j.compstruct.2016.11.046}
  {\path{doi:10.1016/j.compstruct.2016.11.046}}.

\bibitem{banerjee_mechanical_2014}
S.~Banerjee, B.~V. Sankar,
  \href{https://linkinghub.elsevier.com/retrieve/pii/S1359836813006409}{Mechanical
  properties of hybrid composites using finite element method based
  micromechanics}, Composites Part B: Engineering 58 (2014) 318--327.
\newblock \href {https://doi.org/10.1016/j.compositesb.2013.10.065}
  {\path{doi:10.1016/j.compositesb.2013.10.065}}.

\bibitem{devireddy_effect_2014}
S.~B.~R. Devireddy, S.~Biswas,
  \href{https://www.hindawi.com/journals/jcomp/2014/629175/}{Effect of {Fiber}
  {Geometry} and {Representative} {Volume} {Element} on {Elastic} and {Thermal}
  {Properties} of {Unidirectional} {Fiber}-{Reinforced} {Composites}}, Journal
  of Composites 2014 (2014) 1--12.
\newblock \href {https://doi.org/10.1155/2014/629175}
  {\path{doi:10.1155/2014/629175}}.

\bibitem{meyer_bi-logistic_1994}
P.~Meyer,
  \href{https://linkinghub.elsevier.com/retrieve/pii/0040162594900426}{Bi-logistic
  growth}, Technological Forecasting and Social Change 47~(1) (1994) 89--102.
\newblock \href {https://doi.org/10.1016/0040-1625(94)90042-6}
  {\path{doi:10.1016/0040-1625(94)90042-6}}.

\bibitem{reiter_human_2025}
N.~Reiter, S.~Auer, L.~Hoffmann, L.~Bräuer, F.~Paulsen, S.~Budday,
  \href{https://linkinghub.elsevier.com/retrieve/pii/S1742706125002466}{Do
  human brain white matter and brain stem structures show direction-dependent
  mechanical behavior?}, Acta Biomaterialia 199 (2025) 230--251.
\newblock \href {https://doi.org/10.1016/j.actbio.2025.04.004}
  {\path{doi:10.1016/j.actbio.2025.04.004}}.

\bibitem{hou_mechanical_2025}
J.~Hou, K.~Jiang, A.~Ramanathan, A.~S. Kumar, W.~Zhang, L.~Zhao, T.~Wu,
  R.~Pidaparti, D.~Zhu, G.~Li, K.~Song, T.~Liu, M.~J. Razavi, E.~Kuhl, X.~Wang,
  \href{https://arxiv.org/abs/2504.12346}{Mechanical {Characterization} of
  {Brain} {Tissue}: {Experimental} {Techniques}, {Human} {Testing}
  {Considerations}, and {Perspectives}}, version Number: 2 (2025).
\newblock \href {https://doi.org/10.48550/ARXIV.2504.12346}
  {\path{doi:10.48550/ARXIV.2504.12346}}.

\bibitem{bouhrara_agerelated_2021}
M.~Bouhrara, R.~W. Kim, N.~Khattar, W.~Qian, C.~M. Bergeron, D.~Melvin, L.~M.
  Zukley, L.~Ferrucci, S.~M. Resnick, R.~G. Spencer,
  \href{https://onlinelibrary.wiley.com/doi/10.1002/hbm.25372}{Age‐related
  estimates of aggregate \textit{g} ‐ratio of white matter structures
  assessed using quantitative magnetic resonance neuroimaging}, Human Brain
  Mapping 42~(8) (2021) 2362--2373.
\newblock \href {https://doi.org/10.1002/hbm.25372}
  {\path{doi:10.1002/hbm.25372}}.

\bibitem{kang_systematic_2024}
W.~Kang, Q.~Li, L.~Wang, Y.~Zhang, P.~Xu, Y.~Fan,
  \href{https://linkinghub.elsevier.com/retrieve/pii/S2405844024140108}{Systematic
  analysis of constitutive models of brain tissue materials based on
  compression tests}, Heliyon 10~(18) (2024) e37979.
\newblock \href {https://doi.org/10.1016/j.heliyon.2024.e37979}
  {\path{doi:10.1016/j.heliyon.2024.e37979}}.

\bibitem{bernal_mechanical_2007}
R.~Bernal, P.~A. Pullarkat, F.~Melo,
  \href{https://link.aps.org/doi/10.1103/PhysRevLett.99.018301}{Mechanical
  {Properties} of {Axons}}, Phys. Rev. Lett. 99~(1) (2007) 018301.
\newblock \href {https://doi.org/10.1103/PhysRevLett.99.018301}
  {\path{doi:10.1103/PhysRevLett.99.018301}}.

\bibitem{ouyang_contribution_2013}
H.~Ouyang, E.~Nauman, R.~Shi,
  \href{https://jbioleng.biomedcentral.com/articles/10.1186/1754-1611-7-21}{Contribution
  of cytoskeletal elements to the axonal mechanical properties}, J Biol Eng
  7~(1) (2013) 21.
\newblock \href {https://doi.org/10.1186/1754-1611-7-21}
  {\path{doi:10.1186/1754-1611-7-21}}.

\bibitem{atashgar_structure_2025}
F.~Atashgar, M.~Shafieian, N.~Abolfathi,
  \href{https://link.springer.com/10.1007/s10237-025-01957-4}{From structure to
  mechanics: exploring the role of axons and interconnections in anisotropic
  behavior of brain white matter}, Biomech Model Mechanobiol 24~(3) (2025)
  779--810.
\newblock \href {https://doi.org/10.1007/s10237-025-01957-4}
  {\path{doi:10.1007/s10237-025-01957-4}}.

\bibitem{budday_mechanical_2017}
S.~Budday, G.~Sommer, C.~Birkl, C.~Langkammer, J.~Haybaeck, J.~Kohnert,
  M.~Bauer, F.~Paulsen, P.~Steinmann, E.~Kuhl, G.~Holzapfel,
  \href{https://linkinghub.elsevier.com/retrieve/pii/S1742706116305633}{Mechanical
  characterization of human brain tissue}, Acta Biomaterialia 48 (2017)
  319--340.
\newblock \href {https://doi.org/10.1016/j.actbio.2016.10.036}
  {\path{doi:10.1016/j.actbio.2016.10.036}}.

\bibitem{bertalan_mechanical_2023}
G.~Bertalan, J.~Becker, H.~Tzschätzsch, A.~Morr, H.~Herthum, M.~Shahryari,
  R.~D. Greenhalgh, J.~Guo, L.~Schröder, C.~Alzheimer, S.~Budday, K.~Franze,
  J.~Braun, I.~Sack,
  \href{https://linkinghub.elsevier.com/retrieve/pii/S1751616122005185}{Mechanical
  behavior of the hippocampus and corpus callosum: {An} attempt to reconcile ex
  vivo with in vivo and micro with macro properties}, Journal of the Mechanical
  Behavior of Biomedical Materials 138 (2023) 105613.
\newblock \href {https://doi.org/10.1016/j.jmbbm.2022.105613}
  {\path{doi:10.1016/j.jmbbm.2022.105613}}.

\bibitem{guo_towards_2013}
J.~Guo, S.~Hirsch, A.~Fehlner, S.~Papazoglou, M.~Scheel, J.~Braun, I.~Sack,
  \href{https://dx.plos.org/10.1371/journal.pone.0071807}{Towards an
  {Elastographic} {Atlas} of {Brain} {Anatomy}}, PLoS ONE 8~(8) (2013) e71807.
\newblock \href {https://doi.org/10.1371/journal.pone.0071807}
  {\path{doi:10.1371/journal.pone.0071807}}.

\bibitem{morrison_atlas_2023}
O.~Morrison, M.~Destrade, B.~B. Tripathi,
  \href{https://linkinghub.elsevier.com/retrieve/pii/S1742706123004348}{An
  atlas of the heterogeneous viscoelastic brain with local power-law
  attenuation synthesised using {Prony}-series}, Acta Biomaterialia 169 (2023)
  66--87.
\newblock \href {https://doi.org/10.1016/j.actbio.2023.07.040}
  {\path{doi:10.1016/j.actbio.2023.07.040}}.

\bibitem{vestrum2015finite}
O.~Vestrum, Finite element implementation of fibre-reinforced materials model
  in abaqus/explicit, Master's thesis, NTNU (2015).

\bibitem{liu2019dynamic}
P.~Liu, X.~Li, Dynamic void growth and localization behaviors of glassy polymer
  using nonlocal explicit finite element analysis, Journal of Peridynamics and
  Nonlocal Modeling 1 (2019) 3--13.

\bibitem{zheng2024application}
G.~Zheng, N.~Zhang, S.~Lv, The application of fractional derivative
  viscoelastic models in the finite element method: taking several common
  models as examples, Fractal and Fractional 8~(2) (2024) 103.

\bibitem{hasan_viscoelastic_2022}
F.~Hasan, K.~A. Mahmud, M.~I. Khan, A.~Adnan,
  \href{https://www.frontiersin.org/articles/10.3389/fbioe.2022.904818/full}{Viscoelastic
  damage evaluation of the axon}, Front. Bioeng. Biotechnol. 10 (2022) 904818.
\newblock \href {https://doi.org/10.3389/fbioe.2022.904818}
  {\path{doi:10.3389/fbioe.2022.904818}}.

\bibitem{giudice_calibration_2021}
J.~S. Giudice, A.~Alshareef, T.~Wu, A.~K. Knutsen, L.~V. Hiscox, C.~L. Johnson,
  M.~B. Panzer,
  \href{https://www.frontiersin.org/articles/10.3389/fbioe.2021.664268/full}{Calibration
  of a {Heterogeneous} {Brain} {Model} {Using} a {Subject}-{Specific} {Inverse}
  {Finite} {Element} {Approach}}, Front. Bioeng. Biotechnol. 9 (2021) 664268.
\newblock \href {https://doi.org/10.3389/fbioe.2021.664268}
  {\path{doi:10.3389/fbioe.2021.664268}}.

\bibitem{babaei_machine_2022}
H.~Babaei, E.~A. Mendiola, S.~Neelakantan, Q.~Xiang, A.~Vang, R.~A.~F. Dixon,
  D.~J. Shah, P.~Vanderslice, G.~Choudhary, R.~Avazmohammadi,
  \href{https://www.nature.com/articles/s41598-022-09128-6}{A machine learning
  model to estimate myocardial stiffness from {EDPVR}}, Sci Rep 12~(1) (2022)
  5433.
\newblock \href {https://doi.org/10.1038/s41598-022-09128-6}
  {\path{doi:10.1038/s41598-022-09128-6}}.

\bibitem{yuan_fast_2024}
H.~Yuan, X.~Xie, \href{https://arxiv.org/abs/2410.01467}{A fast fully discrete
  mixed finite element scheme for fractional viscoelastic models of wave
  propagation}, version Number: 5 (2024).
\newblock \href {https://doi.org/10.48550/ARXIV.2410.01467}
  {\path{doi:10.48550/ARXIV.2410.01467}}.

\end{thebibliography}

\end{document}